\documentclass[onecolumn]{aastex62}
\usepackage[T1]{fontenc}
\usepackage[english]{babel}
\usepackage[utf8]{inputenc}
\usepackage{amsmath}
\usepackage{amssymb}
\usepackage{latexsym}
\usepackage{cases}
\usepackage{slashbox}
\usepackage{dcolumn}
\usepackage{comment}
\usepackage{multirow}
\usepackage{commath}
\usepackage{bm}
\usepackage{tikz}
\usepackage{amsfonts}
\usepackage{float}
\usepackage{graphicx}
\usepackage{xcolor}
\usetikzlibrary{fadings}

\shorttitle{standing MHD waves in asymmetric magnetic slabs II}
\shortauthors{Oxley, Zs\'{a}mberger \& Erd\'{e}lyi}

\begin{document}

\title{Standing MHD Waves in a Magnetic Slab Embedded in an Asymmetric Magnetic Plasma Environment: \\ Surface Waves}

\correspondingauthor{R\'{o}bert Erd\'{e}lyi}
\email{robertus@sheffield.ac.uk}

\author{William Oxley}
\affiliation{Solar Physics and Space Plasma Research Centre, School of Mathematics and Statistics, University of Sheffield, Hicks Building, Hounsfield Road Sheffield, S3 7RH, UK}

\author{No\'{e}mi Kinga Zs\'{a}mberger}
\affiliation{Solar Physics and Space Plasma Research Centre, School of Mathematics and Statistics, University of Sheffield, Hicks Building, Hounsfield Road Sheffield, S3 7RH, UK}
\affiliation{Department of Physics, University of Debrecen, Egyetem t\'{e}r 1., Debrecen, H-4010, Hungary}
\affiliation{Doctoral School of Physics, University of Debrecen, Egyetem t\'{e}r 1., Debrecen, H-4010, Hungary}

\author{R\'{o}bert Erd\'{e}lyi}
\affiliation{Solar Physics and Space Plasma Research Centre, School of Mathematics and Statistics, University of Sheffield, Hicks Building, Hounsfield Road Sheffield, S3 7RH, UK}
\affiliation{Department of Astronomy, E\"{o}tv\"{o}s Lor\'{a}nd University, 1/A P\'{a}zm\'{a}ny P\'{e}ter s\'{e}t\'{a}ny, H-1117, Budapest, Hungary}

\begin{abstract}
Building on a previous study that analyses surface waves in magnetic slabs embedded in a non-magnetic external environment, in this study the model is generalised and external magnetic fields are added. The slab is assumed to be thin, with weak magnetic asymmetry. The frequencies of the standing harmonic modes are derived to leading-order in the small quantities representing the thin slab width and the weak asymmetry. It is found that the frequencies are more prone to changes to the width of the slab than changes in the magnetic asymmetry. The frequency ratio of the first harmonic to the fundamental mode is derived, along with the amplitude difference between the two sides of the slab, as these may be observable quantities that can be compared with observational results and applied to carry out solar magneto-seismology.

\end{abstract}

\section{Introduction} \label{introduction}
Many of the magnetic structures in the solar atmosphere have been observed to support oscillations on a variety of scales from coronal loops (\citealt{Aschwanden-99}; \citealt{Wang-04}; \citealt{Ban-07}; \citealt{demoor-09}) and prominences \citep{Arregui-12} to spicules (\citealt{zaq-09}; \citealt{tsir-12}), sunspot light bridges (\citealt{yuan-14}; \citealt{yang-16, yang-17}) and magnetic bright points \citep{Liu-18}. An analytical wave study using the magnetohydrodynamic approximation can lead to the application of solar magneto-seismology (SMS; see reviews by \citealt{nak-05}; \citealt{erd-06-a, erd-06-b}; \citealt{andries-09}; \citealt{rud-erd-09}), which makes it possible to deduce values for background parameters that are not directly measurable.

A popular mathematical framework used to describe these oscillations was introduced by \cite{rob-81-a}, where wave propagation at a single interface (more precisely, a discontinuity in the system parameters such as temperature or magnetic field strength), was studied. This was shortly followed by another investigation on wave propagation in a magnetic slab embedded in a non-magnetic symmetric external environment \citep{rob-81-b}, and then further generalised by considering a magnetic external environment \citep{rob-82}. This mathematical model is extremely powerful as it is still being used nearly two full solar cycles after its introduction. More recent studies have further generalised this framework, where cylindrical instead of Cartesian geometry is used and the flux tube is studied as opposed to the magnetic slab (e.g. \citealt{ab-88}). In addition to this, work has been further expanded with the examination of MHD waves within an asymmetric slab system, with both a non-magnetic \citep{all-erd-17} and a magnetic \citep{zsam-all-18} external environment. A further generalisation of this model was provided by \cite{shuk-18}, who analysed multiple slab systems placed side by side to create the n-slab model. Some important applications of these recent analytical modeling studies were presented by \cite{all-19}, which included the amplitude ratio method as an SMS technique.

All these previous analytical approaches investigated propagating MHD waves. However, observations have discovered a large range of standing waves in MHD waveguides \citep{demoor-09}. A review by \cite{tar-09} presents recent findings relating to MHD waves in the context of coronal heating. In particular, there is some discussion regarding standing waves and the possibility of using SMS to deduce diagnostic information, with the particular example of coronal loops included. Other studies of standing waves in coronal loops include \cite{nak-01}, \cite{wang-07} and \cite{wang-11}. These investigations also provide examples where an estimation of the magnetic field strength is obtained using standing MHD oscillations. This illustrates how an analysis of standing MHD waves can lead to the application of SMS, and demonstrates the power of this technique by showing how it may be possible to derive diagnostic information about the waveguide. Investigations of standing oscillations in other solar features have been conducted, such as a solar prominence, which is in fact an example of a solar structure that is well suited to the magnetic slab model (\citealt{ol-09}; \citealt{Arregui-12}), and consequently further understanding standing modes in a magnetic slab could assist in understanding the behaviour of phenomena such as the solar prominence. Therefore, \cite{ox-20} embarked on investigating standing modes in a magnetic slab embedded in a non-magnetic asymmetric environment.

The current paper now aims to build on \cite{ox-20} by introducing magnetism to the external plasma environment. By assuming both the thin slab and weak asymmetry approximations, the standing waves will be examined by deriving the dispersion equations for the frequencies and amplitudes in terms of small parameters that represent thin slab width and weak magnetic asymmetry. Introducing these small parameters allows analytical progress to be made while still keeping some complex aspects of an asymmetric environment. The external region of plasma will be assumed to be isothermal, meaning there is no asymmetry in the temperature and the effects of asymmetry in the external magnetic fields can be isolated and examined. Some of the derived quantities that relate to the frequencies and amplitudes of oscillation have analogous results in the study of a magnetic slab embedded in a non-magnetic asymmetric environment \citep{ox-20}, and similarities will be discussed throughout.

The main purpose of this study is to further advance the magnetic slab model, allowing solar magneto-seismology to be used for a wider range of suitable solar structures. More precisely, by carrying out the analytical investigation, quantities that depend on magnetic asymmetry (such as the amplitude difference between the two sides of the slab) are derived, and it may be possible to infer values of the internal or external magnetic field strength, which are not directly observable.

Standing modes in a solar prominence were investigated in \cite{joarder}, using a magnetic slab model. In this study the analytical work was very short, and in fact contained an error, as indicated in \cite{ox-20}. The present study, along with \cite{ox-20}, provides a more detailed and accurate analysis of standing modes in a magnetic slab structure.

First, we present the equilibrium of the model along with the appropriate line-tying boundary conditions and pressure balance requirements (which are necessary in order to have a static waveguide with no background bulk motions). Then, the general dispersion relation that governs the surface and body waves of the slab is derived in Section \ref{The General Dispersion Relation}. This is done in order to identify what frequencies of waves the slab will support, and then these findings can be used to derive potentially observable quantities. In Section \ref{Reduction of the Dispersion Relation}, the thin slab and weak asymmetry assumptions are applied in order to have an analytical insight into the problem. A magneto-seismology examination of the frequencies of the standing harmonic modes will be presented in Section \ref{Frequencies of the Standing Harmonic Modes}, by deriving analytic expressions for these eigenfrequencies in terms of parameters representing slab width and magnetic asymmetry.  Analysis of the amplitudes of the eigenmodes is given in Section \ref{Amplitudes of the Standing Harmonic Modes}. We conclude with a short discussion of the results.

\section{The Equilibrium Magnetic Slab} \label{The Equilibrium Magnetic Slab}

\begin{figure}[h!]
\centering
\begin{tikzpicture}

%fill middle
\path [fill=lightgray, opacity=0.7] (3.5,1) -- (3.5,5) -- (6.5,5) -- (6.5,1) -- (3.5,1);
\shade[left color=lightgray,right color=white, opacity=0.7] (6.5,1) -- (6.5,5) -- (7,5.3) -- (7,1.3) -- (6.5,1);
\shade[top color=lightgray,bottom color=white, opacity=0.7] (3.5,1) -- (6.5,1) -- (6.5,0.7) -- (3.5,0.7) -- (3.5,1);
\shade[top color=white,bottom color=lightgray, opacity=0.7] (3.5,5) -- (4,5.3) -- (7,5.3) -- (6.5,5) -- (3.5,5);
\shade[left color=lightgray,right color=white, opacity=0.7] (6.5,1) -- (6.5,0.7) -- (7,1) -- (7,1.3) -- (6.5,1);

%fill left
\path [fill=lightgray, opacity=0.4] (-0.5,1) -- (-1,2) -- (-0.5,3) -- (-1,4) -- (-0.5,5) -- (3.5,5) -- (3.5,1) -- (0,1);
\shade[top color=lightgray,bottom color=white, opacity=0.45] (0,1) -- (3.5,1) -- (3.5,0.7) -- (0,0.7) -- (0,1);
\shade[top color=white,bottom color=lightgray, opacity=0.45] (0,5) -- (0.5,5.3) -- (4,5.3) -- (3.5,5) -- (0,5);
\shade[top color=lightgray, bottom color=lightgray, opacity=0.3] (-0.5,1.) to (-0.5,0.) to (3.5,0) to (3.5,1) to (-0.5,1);
\shade[top color=lightgray, bottom color=lightgray, opacity=0.3] (-0.5,5.) to (-0.5,6.) to (3.5,6) to (3.5,5) to (-0.5,5);

%middle top
\path [fill=lightgray, opacity=0.6] (3.5,5) -- (3.5,6) -- (6.5,6) -- (6.5,5) -- (3.5,5);
\path [fill=lightgray, opacity=0.6] (3.5,0) -- (3.5,1) -- (6.5,1) -- (6.5,0) -- (3.5,0);
\shade[top color=lightgray,bottom color=white, opacity=0.45] (3.5,0) -- (6.5,0) -- (6.5,-0.3) -- (3.5,-0.3) -- (3.5,0);
\shade[top color=white,bottom color=lightgray, opacity=0.45] (3.5,6) -- (4,6.3) -- (7,6.3) -- (6.5,6) -- (3.5,6);
\shade[top color=white,bottom color=lightgray, opacity=0.45] (3.5,5) -- (4,5.3) -- (7,5.3) -- (6.5,5) -- (3.5,5);
\shade[left color=lightgray,right color=white, opacity=0.99] (6.5,0) -- (6.5,-0.3) -- (7,0) -- (7,0.3) -- (6.5,0);

%right top and bottom
\shade[left color=lightgray,right color=white, opacity=0.9] (10,1) -- (9.5,2) -- (10,3) -- (9.5,4) -- (10,5) -- (10.5,5.3) -- (10,4.3) -- (10.5,3.3) -- (10,2.3) -- (10.5,1.3) -- (10,1);
\shade[top color=lightgray, bottom color=lightgray, opacity=0.9] (6.5,1.) to (6.5,0.) to (10.,0) to (10.,1) to (6.5,1);
\shade[top color=lightgray, bottom color=lightgray, opacity=0.9] (6.5,5.) to (6.5,6.) to (10.,6) to (10.,5) to (6.5,5);

%top and bottom corners right
\shade[top color=white,bottom color=black, opacity=0.9] (6.5,1) -- (7.,1.3) -- (10.5,1.3) -- (10,1) -- (6.5,1);
\shade[top color=white,bottom color=black, opacity=0.9] (6.5,5) -- (7.,5.3) -- (10.5,5.3) -- (10,5) -- (6.5,5);

%surface on the left
\draw [ultra thick, dashed] (3.5,0) -- (3.5,6);
\draw [ultra thick, dashed, path fading=east] (3.5,0) -- (4,0.3);
\draw [ultra thick, dashed, path fading=east] (3.5,4) -- (4,4.3);
\draw [ultra thick, dashed, path fading=east] (3.5,2) -- (4,2.3);
\draw [ultra thick, dashed, path fading=east] (3.5,1) -- (4,1.3);
\draw [ultra thick, dashed, path fading=east] (3.5,3) -- (4,3.3);
\draw [ultra thick, dashed, path fading=east] (3.5,4) -- (4,4.3);
\draw [ultra thick, dashed, path fading=east] (3.5,5) -- (4,5.3);
\draw [ultra thick, dashed, path fading=east] (3.5,6) -- (4,6.3);

%outer surface left
\shade[top color=lightgray,bottom color=white, opacity=0.3] (-0.5,0) -- (3.5,0) -- (3.5,-0.3) -- (-0.5,-0.3) -- (-0.5,0);
\shade[top color=white,bottom color=lightgray, opacity=0.3] (-0.5,6) -- (0.,6.3) -- (4,6.3) -- (3.5,6) -- (-0.5,6);
\shade[top color=white,bottom color=lightgray, opacity=0.7] (-0.5,5) -- (0.,5.3) -- (4,5.3) -- (3.5,5) -- (-0.5,5);
\shade[right color=white,left color=lightgray, opacity=0.5] (0,1.3) to (-0.5,2.3) to  (0,3.3) to (-0.5,4.3) to (0,5.3) to (0,5.3) to (0,3.3) to  (0,1.3) to (-0.5,1.3);

%fill right
\path [fill=lightgray, opacity=0.9] (6.5,1) -- (6.5,5) -- (10,5) -- (9.5,4) -- (10,3) -- (9.5,2) -- (10,1) -- (6.5,1);
\shade[top color=lightgray,bottom color=white, opacity=0.8] (6.5,1) -- (10,1) -- (10,0.7) -- (6.5,0.7) -- (6.5,1);
\shade[top color=white,bottom color=lightgray, opacity=0.8] (6.5,5) -- (7,5.3) -- (10.5,5.3) -- (10,5) -- (6.5,5);
\shade[top color=lightgray, bottom color=lightgray, opacity=0.75] (6.5,1.) to (6.5,0.) to (10,0) to (10,1) to (6.5,1);
\shade[top color=lightgray, bottom color=lightgray, opacity=0.75] (6.5,5.) to (6.5,6.) to (10,6) to (10,5) to (6.5,5);

%%surface on the right
%\shade[left color=darkgray,right color=white, opacity=0.99] (6.5,0) -- (6.5,1) -- (7,1.3) -- (7,0.3) -- (6.5,0);
%\shade[left color=darkgray,right color=white, opacity=0.99] (6.5,5) -- (6.5,6) -- (7,6.3) -- (7,5.3) -- (6.5,5);

%surface on the right
\draw [ultra thick, dashed] (6.5,0) --(6.5,6);
\draw [ultra thick, dashed, path fading=east] (6.5,0) -- (7,0.3);
\draw [ultra thick, dashed, path fading=east] (6.5,4) -- (7,4.3);
\draw [ultra thick, dashed, path fading=east] (6.5,2) -- (7,2.3);
\draw [ultra thick, dashed, path fading=east] (6.5,1) -- (7,1.3);
\draw [ultra thick, dashed, path fading=east] (6.5,3) -- (7,3.3);
\draw [ultra thick, dashed, path fading=east] (6.5,4) -- (7,4.3);
\draw [ultra thick, dashed, path fading=east] (6.5,5) -- (7,5.3);
\draw [ultra thick, dashed, path fading=east] (6.5,6) -- (7,6.3);

%left bottom
\shade[top color=lightgray,bottom color=white, opacity=0.3] (-0.5,0) -- (3.5,0) -- (3.5,-0.3) -- (-0.5,-0.3) -- (-0.5,0);
\shade[top color=white,bottom color=lightgray, opacity=0.7] (-0.5,1) -- (0.,1.3) -- (4,1.3) -- (3.5,1) -- (-0.5,1);
\shade[bottom color=lightgray,top color=white, opacity=0.45] (3.5,1) -- (6.5,1) -- (6.5,1.3) -- (4.,1.3) -- (3.5,1);

%left top dashed
\draw [ultra thick, dashed] (-0.5,5.) -- (3.5,5);
\draw [ultra thick, dashed, path fading = east] (-0.5,5.) -- (0,5.3);
\draw [ultra thick, dashed, path fading = east] (0.5,5.) -- (1,5.3);
\draw [ultra thick, dashed, path fading = east] (1.5,5.) -- (2,5.3);
\draw [ultra thick, dashed, path fading = east] (2.5,5.) -- (3,5.3);

%left bottom dashed
\draw [ultra thick, dashed] (-0.5,1.) -- (3.5,1);
\draw [ultra thick, dashed, path fading = east] (-0.5,1.) -- (0,1.3);
\draw [ultra thick, dashed, path fading = east] (0.5,1.) -- (1,1.3);
\draw [ultra thick, dashed, path fading = east] (1.5,1.) -- (2,1.3);
\draw [ultra thick, dashed, path fading = east] (2.5,1.) -- (3,1.3);

%right top dashed
\draw [ultra thick, dashed] (3.5,5.) -- (3.5,6);
\draw [ultra thick, dashed] (6.5,5) --(10,5);
\draw [ultra thick, dashed, path fading=east] (7.5,5) -- (8,5.3);
\draw [ultra thick, dashed, path fading=east] (8.5,5) -- (9,5.3);
\draw [ultra thick, dashed, path fading=east] (9.5,5) -- (10,5.3);

%right bottom dashed
\draw [ultra thick, dashed] (6.5,1) --(10,1);
\draw [ultra thick, dashed, path fading=east] (7.5,1) -- (8,1.3);
\draw [ultra thick, dashed, path fading=east] (8.5,1) -- (9,1.3);
\draw [ultra thick, dashed, path fading=east] (9.5,1) -- (10,1.3);

%top and bottom dashed middle
\draw [ultra thick, dashed] (3.5,5) --(6.5,5);
\draw [ultra thick, dashed, path fading=east] (4.5,5) -- (5,5.3);
\draw [ultra thick, dashed, path fading=east] (5.5,5) -- (6,5.3);
\draw [ultra thick, dashed] (3.5,1) --(6.5,1);
\draw [ultra thick, dashed, path fading=east] (4.5,1) -- (5,1.3);
\draw [ultra thick, dashed, path fading=east] (5.5,1) -- (6,1.3);

%right
\shade[top color=lightgray,bottom color=white, opacity=0.8] (6.5,0) -- (10,0) -- (10,-0.3) -- (6.5,-0.3) -- (6.5,0);
\shade[top color=white,bottom color=lightgray, opacity=0.8] (6.5,6) -- (7,6.3) -- (10.5,6.3) -- (10,6) -- (6.5,6);
\shade[right color=white,left color=lightgray, opacity=0.8] (10,-0.) -- (10,1) -- (10.5,1.3) -- (10.5,0.3) -- (10,-0.);
\shade[bottom color=white,top color=lightgray, opacity=0.3] (10,-0.) -- (10,1) -- (10.5,1.3) -- (10.5,0.3) -- (10,-0.);
\shade[right color=white, left color=lightgray, opacity=0.8] (10,5.) -- (10,6) -- (10.5,6.3) -- (10.5,5.3) -- (10,5.);
\shade[top color=white, bottom color=lightgray, opacity=0.8] (10,5.) -- (10,6) -- (10.5,6.3) -- (10.5,5.3) -- (10,5.);
\shade[top color=lightgray,bottom color=white, opacity=0.3] (10,-0.3) -- (10,0.) -- (10.5,0.3) -- (10.5,-0.) -- (10,-0.3);
\shade[left color=lightgray, right color=white, opacity=0.3] (10,-0.3) -- (10,0.) -- (10.5,0.3) -- (10.5,-0.) -- (10,-0.3);

%right top and bottom
\path [fill=lightgray, opacity=0.7] (6.5,5) -- (6.5,6) -- (10,6) -- (10,5) -- (6.5,5);
\path [fill=lightgray, opacity=0.9] (6.5,0) -- (6.5,1) -- (10,1) -- (10,0) -- (6.5,0);

%0
%magnetic field lines
\draw [ultra thick, red, path fading=north] (4.0,1) -- (4.0, 2.7);
\draw [ultra thick, red, path fading=south] (4.0,3.2) -- (4.0,5);
\draw [ultra thick, red, -stealth] (4,4.15) -- (4,4.25);
\draw [ultra thick, red, -stealth] (4,1.75) -- (4,1.85);
\draw [ultra thick, red, path fading=north] (4.5,1) -- (4.5, 2.7); 
\draw [ultra thick, red, path fading=south] (4.5,3.2) -- (4.5,5);
\draw [ultra thick, red, -stealth] (4.5,4.15) -- (4.5,4.25);
\draw [ultra thick, red, -stealth] (4.5,1.75) -- (4.5,1.85);
\draw [ultra thick, red, path fading=north] (5,1) -- (5, 2.7);
\draw [ultra thick, red, path fading=south] (5,3.2) -- (5,5);
\draw [ultra thick, red, -stealth] (5,4.15) -- (5,4.25);
\draw [ultra thick, red, -stealth] (5,1.75) -- (5,1.85);
\draw [ultra thick, red, path fading=north] (5.5,1) -- (5.5, 2.7);
\draw [ultra thick, red, path fading=south] (5.5,3.2) -- (5.5,5);
\draw [ultra thick, red, -stealth] (5.5,4.15) -- (5.5,4.25);
\draw [ultra thick, red, -stealth] (5.5,1.75) -- (5.5,1.85);
\draw [ultra thick, red, path fading=north] (6.0,1) -- (6.0, 2.7);
\draw [ultra thick, red, path fading=south] (6.0,3.2) -- (6.0,5);
\draw [ultra thick, red, -stealth] (6,4.15) -- (6,4.25);
\draw [ultra thick, red, -stealth] (6,1.75) -- (6,1.85);

%magnetic field lines right
\draw [ultra thick, red, path fading=north] (6.9,1) -- (6.9, 2.7);
\draw [ultra thick, red, path fading=south] (6.9,3.2) -- (6.9,5);
\draw [ultra thick, red, -stealth] (6.9,4.15) -- (6.9,4.25);
\draw [ultra thick, red, -stealth] (6.9,1.75) -- (6.9,1.85);
\draw [ultra thick, red, path fading=north] (7.2,1) -- (7.2, 2.7);
\draw [ultra thick, red, path fading=south] (7.2,3.2) -- (7.2,5);
\draw [ultra thick, red, -stealth] (7.2,4.15) -- (7.2,4.25);
\draw [ultra thick, red, -stealth] (7.2,1.75) -- (7.2,1.85);
\draw [ultra thick, red, path fading=north] (7.5,1) -- (7.5, 2.7); 
\draw [ultra thick, red, path fading=south] (7.5,3.2) -- (7.5,5);
\draw [ultra thick, red, -stealth] (7.5,4.15) -- (7.5,4.25);
\draw [ultra thick, red, -stealth] (7.5,1.75) -- (7.5,1.85);
\draw [ultra thick, red, path fading=north] (7.8,1) -- (7.8, 2.7);
\draw [ultra thick, red, path fading=south] (7.8,3.2) -- (7.8,5);
\draw [ultra thick, red, -stealth] (7.8,4.15) -- (7.8,4.25);
\draw [ultra thick, red, -stealth] (7.8,1.75) -- (7.8,1.85);
\draw [ultra thick, red, path fading=north] (8.1,1) -- (8.1, 2.7);
\draw [ultra thick, red, path fading=south] (8.1,3.2) -- (8.1,5);
\draw [ultra thick, red, -stealth] (8.1,4.15) -- (8.1,4.25);
\draw [ultra thick, red, -stealth] (8.1,1.75) -- (8.1,1.85);
\draw [ultra thick, red, path fading=north] (8.4,1) -- (8.4, 2.7);
\draw [ultra thick, red, path fading=south] (8.4,3.2) -- (8.4,5);
\draw [ultra thick, red, -stealth] (8.4,4.15) -- (8.4,4.25);
\draw [ultra thick, red, -stealth] (8.4,1.75) -- (8.4,1.85);
\draw [ultra thick, red, path fading=north] (8.7,1) -- (8.7, 2.7);
\draw [ultra thick, red, path fading=south] (8.7,3.2) -- (8.7,5);
\draw [ultra thick, red, -stealth] (8.7,4.15) -- (8.7,4.25);
\draw [ultra thick, red, -stealth] (8.7,1.75) -- (8.7,1.85);
\draw [ultra thick, red, path fading=north] (9,1) -- (9, 2.7); 
\draw [ultra thick, red, path fading=south] (9,3.2) -- (9,5);
\draw [ultra thick, red, -stealth] (9,4.15) -- (9,4.25);
\draw [ultra thick, red, -stealth] (9,1.75) -- (9,1.85);
\draw [ultra thick, red, path fading=north] (9.3,1) -- (9.3, 2.7);
\draw [ultra thick, red, path fading=south] (9.3,3.2) -- (9.3,5);
\draw [ultra thick, red, -stealth] (9.3,4.15) -- (9.3,4.25);
\draw [ultra thick, red, -stealth] (9.3,1.75) -- (9.3,1.85);

%%magnetic field lines left
\draw [ultra thick, red, path fading=north] (0.2,1) -- (0.2, 2.7);
\draw [ultra thick, red, path fading=south] (0.2,3.2) -- (0.2,5);
\draw [ultra thick, red, -stealth] (0.2,4.15) -- (0.2,4.25);
\draw [ultra thick, red, -stealth] (0.2,1.75) -- (0.2,1.85);
\draw [ultra thick, red, path fading=north] (1.2,1) -- (1.2, 2.7); 
\draw [ultra thick, red, path fading=south] (1.2,3.2) -- (1.2,5);
\draw [ultra thick, red, -stealth] (1.2,4.15) -- (1.2,4.25);
\draw [ultra thick, red, -stealth] (1.2,1.75) -- (1.2,1.85);
\draw [ultra thick, red, path fading=north] (2.2,1) -- (2.2, 2.7);
\draw [ultra thick, red, path fading=south] (2.2,3.2) -- (2.2,5);
\draw [ultra thick, red, -stealth] (2.2,4.15) -- (2.2,4.25);
\draw [ultra thick, red, -stealth] (2.2,1.75) -- (2.2,1.85);
\draw [ultra thick, red, path fading=north] (3.2,1) -- (3.2, 2.7);
\draw [ultra thick, red, path fading=south] (3.2,3.2) -- (3.2,5);
\draw [ultra thick, red, -stealth] (3.2,4.15) -- (3.2,4.25);
\draw [ultra thick, red, -stealth] (3.2,1.75) -- (3.2,1.85);

\draw [<-] (0,6) -- (0,-0.3); 	%z
\draw [->] (0,1) -- (10,1);		%x
\draw [->] (0,1) -- (0.5,1.3);	%y

\small
\node [below] at (3.5,-0.2) {$x=-x_0$};
\node [below] at (6.5,-0.2) {$x=x_0$};
\node [below] at (10,1) {$x$};
\node [left] at (0,6) {$z$};
\node [right] at (0.45,1.3) {$y$};
\node [left] at (-0.5,1) {$z=0$};
\node [left] at (-0.5,5) {$z=L$};

\large
\node [right] at (0.35,3) {$\rho_1$, $p_1$, $T_1$, $B_1$};
\node [right] at (3.75,3) {$\rho_0$, $p_0$, $T_0$, $B_0$};
\node [right] at (7.1,3) {$\rho_2$, $p_2$, $T_2$, $B_2$};

%height
\draw [very thick] (3.3,4.9) -- (3.3,3.2);
\draw [very thick] (3.3,2.8) -- (3.3,1.1);
\draw [very thick] (3.2,4.9) -- (3.4,4.9);
\draw [very thick] (3.2,1.1) -- (3.4,1.1);
\node at (3.3,3) {$L$};

%width
\draw [very thick] (3.6,0.75) -- (4.6,0.75);
\draw [very thick] (5.3,0.75) -- (6.4,0.75);
\draw [very thick] (3.6,0.9) -- (3.6,0.6);
\draw [very thick] (6.4,0.9) -- (6.4,0.6);
\node at (5,0.7) {$2x_0$};

%magnetic field end points left
\fill[red] (0.2,1) circle (0.08cm);
\fill[red] (1.2,1) circle (0.08cm);
\fill[red] (2.2,1) circle (0.08cm);
\fill[red] (3.2,1) circle (0.08cm);
\fill[red] (0.2,5) circle (0.08cm);
\fill[red] (1.2,5) circle (0.08cm);
\fill[red] (2.2,5) circle (0.08cm);
\fill[red] (3.2,5) circle (0.08cm);

%magnetic field end points middle
\fill[red] (4,1) circle (0.08cm);
\fill[red] (4.5,1) circle (0.08cm);
\fill[red] (5,1) circle (0.08cm);
\fill[red] (5.5,1) circle (0.08cm);
\fill[red] (6,1) circle (0.08cm);
\fill[red] (4,5) circle (0.08cm);
\fill[red] (4.5,5) circle (0.08cm);
\fill[red] (5,5) circle (0.08cm);
\fill[red] (5.5,5) circle (0.08cm);
\fill[red] (6,5) circle (0.08cm);

%magnetic field end points right
\fill[red] (6.9,1) circle (0.08cm);
\fill[red] (7.2,1) circle (0.08cm);
\fill[red] (7.5,1) circle (0.08cm);
\fill[red] (7.8,1) circle (0.08cm);
\fill[red] (8.1,1) circle (0.08cm);
\fill[red] (8.4,1) circle (0.08cm);
\fill[red] (8.7,1) circle (0.08cm);
\fill[red] (9,1) circle (0.08cm);
\fill[red] (9.3,1) circle (0.08cm);
\fill[red] (6.9,5) circle (0.08cm);
\fill[red] (7.2,5) circle (0.08cm);
\fill[red] (7.5,5) circle (0.08cm);
\fill[red] (7.8,5) circle (0.08cm);
\fill[red] (8.1,5) circle (0.08cm);
\fill[red] (8.4,5) circle (0.08cm);
\fill[red] (8.7,5) circle (0.08cm);
\fill[red] (9,5) circle (0.08cm);
\fill[red] (9.3,5) circle (0.08cm);

\end{tikzpicture}
\caption{Visualisation of the equilibrium state inside ( $\abs{x}\leq x_0$) and outside of the magnetic slab ($x<-x_0$ and $x>x_0$), where the magnetic field is indicated by the red lines. The two interfaces marked by the dashed lines at $x=-x_0$, $x=x_0$ outline the slab, while the dashed lines at $z=0$ and $z=L$ mark that the slab is bounded.}
\label{fig1}
\end{figure}
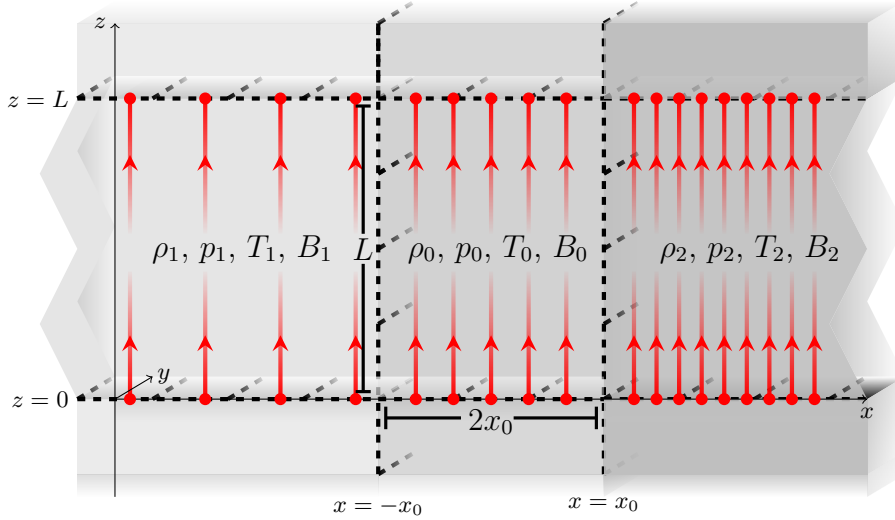

Consider a 3-dimensional, inviscid, static, ideal plasma split into three regions in the $x$-direction, with the equilibrium configuration shown in Figure \ref{fig1}. We assume that the slab is unbounded in the $y$-direction. There is an equilibrium magnetic field given by $B(x)\hat{\bm{e}}_z$, where
\begin{numcases}
{B(x) =} B_1 & \mbox{if $ x < -x_0$}, \\ B_0 & \mbox{if $-x_0 \leq x\leq x_0$}\label{slab1}, \\ B_2 & \mbox{if $ x> x_0$}\label{slab1},
\end{numcases}
and $B_i$, for $i=0,1,2$, are constant. Here, $p_i$, $T_i$ and $\rho_i$ denote the equilibrium kinetic plasma pressure, temperature and density, respectively, where $i=0$ is for the inside of the slab, $i=1$ is for the domain to the left and $i=2$ is for the domain to the right of the slab. The effect of gravity is ignored throughout for simplicity, which may limit the range of application of the results. Also, this work aims to analyse the effects of magnetic asymmetry on the standing surface waves, and the magnetoacoustic-gravity wave supported by the waveguide will not be examined.

\subsection{Boundary Conditions}
The following boundary conditions must be applied at the interfaces, $x=\pm x_0$, and at the end points of the slab, $z=0,L$.

Total pressure balance across the interfaces at $x=\pm x_0$ is required in order for the equilibrium to be stable:

\begin{equation}
p_1+\frac{B_1^2}{2\mu_0}=p_0+\frac{B_0^2}{2\mu_0}=p_2+\frac{B_2^2}{2\mu_0},
\label{slab2}
\end{equation}
where $\mu_0$ is the permeability of free space. We denote the sound speeds by $c_i=\sqrt{\dfrac{\gamma p_i}{\rho_i}}$, and the Alfv\'{e}n speeds by  $v_{Ai}=\dfrac{B_i}{\sqrt{\rho_i\mu_0}}$, for $i=0,1,2$, where $\gamma$ is the adiabatic gas index, which is taken to be constant across the entire system, under the assumption that the plasma composition is the same throughout.

We assume line-tying, which gives boundary conditions that we must apply at $z=0,L$ for the formation of standing waves. These conditions are

\begin{equation}
v_x(z=0)=v_x(z=L)=0, \quad b_z(z=0)=b_z(z=L)=0, \quad p_T(z=0)=p_T(z=L)=0,
\label{slab3}
\end{equation}
where $v_x$ and $b_z$ are the $x$-component of the velocity perturbation and the $z$-component of the magnetic field perturbation, respectively. Additionally, the total pressure perturbation $p_T$ is the sum of the kinetic plasma pressure perturbation and the magnetic pressure perturbation. Using the physical assumption of line-tying, meaning the plasma does not move at $z=0,L$, the condition on $v_x$ follows trivially. The 'frozen-in' property of the field lines implies that the magnetic flux through an arbitrary flux bundle must remain constant. There is no motion at $z=0,L$ by the assumption of line-tying, meaning the cross sectional area of a flux bundle does not change, and consequently the $z$-component of the magnetic field strength remains constant, and the condition on $b_z$ is deduced. The condition on $p_T$ is acquired simply because a non-zero pressure perturbation would cause a force, and consequently cause motion.

\section{Derivation of the Dispersion Relation}  \label{Derivation of the Dispersion Relation}
In this section, the dispersion relation governing the waves in the magnetic slab embedded in an asymmetric magnetic environment will be derived, in both the general case and under the limits of both the thin slab and weak asymmetry approximations.

\subsection{The General Dispersion Relation} \label{The General Dispersion Relation}

Following the derivation given in \cite{rob-81-b}, \cite{ox-20}, and assuming that the velocity perturbation in the $y$- direction, $v_y$, is zero, and that all quantities are independent of $y$, we arrive at two equations for the velocity perturbations in the $x$- and $z$-directions, $v_x$ and $v_z$:

\begin{equation}
\frac{\partial^2v_x}{\partial t^2}=c_0^2\frac{\partial}{\partial x}\left( \boldsymbol\nabla \boldsymbol. \boldsymbol v \right) + v_A^2\boldsymbol\nabla^2v_x,
\label{dr1}
\end{equation}

\begin{equation}
\frac{\partial^2v_z}{\partial t^2}=c_0^2\frac{\partial}{\partial z}\left( \boldsymbol\nabla \boldsymbol. \boldsymbol v \right).
\label{dr2}
\end{equation}
Equations (\ref{dr1}) and (\ref{dr2}) govern the disturbances inside the slab, and to proceed further we must apply the boundary conditions.

\subsubsection{Boundary Conditions}

To satisfy the line-tying boundary condition given by Equation (\ref{slab3}), we assume
\begin{equation} 
v_x =\hat{v}_x(x)e^{-i\omega t}\sin{(kz)}, \quad v_z =\hat{v}_z(x)e^{-i\omega t}f(z), \quad p_T =\hat{p}_T(x)e^{-i\omega t}\sin{(kz)},
\label{dr3}
\end{equation}
where $f$ is an arbitrary function to be determined. The line-tying boundary condition, $v_x(z=0)=v_x(z=L)=0$, gives a restriction on $k$:

\begin{equation}
k=\frac{n\pi}{L}, \quad n \in \mathbb{Z}^+,
\label{dr4}
\end{equation}
where we exclude $k=0$ to have non-trivial solutions.
Using Equations (\ref{dr1}) - (\ref{dr3}), it is straightforward to show that $f(z)=Q\cos(kz)$, for an arbitrary constant $Q$. The constant $Q$ can be absorbed into $\hat{v}_z(x)$, meaning that we can then take $f(z)=\cos(kz)$. 

\subsubsection{The Governing Equations}

Substituting the expressions for $v_x$ and $v_z$ given by Equation (\ref{dr3}) into Equations (\ref{dr1}) and (\ref{dr2}) yield

\begin{equation}
\frac{d^2\hat{v}_x}{dx^2} - m_i^2\hat{v}_x=0, \quad \mbox{where} \quad m_i^2=\frac{(k^2c_i^2-\omega^2)(k^2v_{Ai}^2-\omega^2)}{(k^2c_{Ti}^2-\omega^2)(c_i^2+v_{Ai}^2)}, \quad c_{Ti}^2=\frac{c_i^2v_{Ai}^2}{c_i^2+v_{Ai}^2}.
\label{dr5}
\end{equation}
This governing equation applies to the magnetic plasma environment left of the slab when $i=1$, to the right when $i=2$ and to the magnetic slab in the middle when $i=0$. We assume that $m_0^2>0$ to examine surface waves only, and also assume that $m_1^2,m_2^2>0$ to ensure the solutions are evanescent outside of the slab.

\subsubsection{The Dispersion Relation} \label{DR}

Using Equation (\ref{dr5}) we can write down the solution for $\hat{v}_x$ for all $x$, and using this, it is possible to calculate the total pressure, $p_T$. Applying the continuity of $p_T$ and $v_x$ across both $x=\pm x_0$ gives the dispersion relation for MHD waves in a slab embedded in an asymmetric magnetic environment (with the details given in \cite{zsam-all-18}):

\begin{multline}
2m_0^2(k^2v_{A1}^2-\omega^2)(k^2v_{A2}^2-\omega^2) +2\frac{\rho_0}{\rho_1}m_1\frac{\rho_0}{\rho_2}m_2\left(k^2v_{A0}^2-\omega^2\right)^2 \\ + m_0\left(k^2v_{A0}^2-\omega^2\right)\left(\frac{\rho_0}{\rho_1}m_1(k^2v_{A2}^2-\omega^2) + \frac{\rho_0}{\rho_2}m_2(k^2v_{A1}^2-\omega^2)\right)\left(\tanh{(m_0x_0)}+\coth{(m_0x_0)}\right) = 0. \\
\label{dr6}
\end{multline}
If the asymmetry is removed, after some algebra, one recovers Equation (11) in \cite{rob-82}, the dispersion relation for MHD waves in the magnetic slab embedded in a magnetically symmetric environment.

\subsection{Reduction of the Dispersion Relation} \label{Reduction of the Dispersion Relation}

This section is focused on simplifying the dispersion relation under various assumptions that have practical relevance, e.g. to apply solar magnetoseismology. Firstly, the external regions are assumed to be isothermal in order to emphasize the effects of the magnetic asymmetry. The assumption manifests in the following relations:

\begin{equation}
T_1=T_2, \quad c_1=c_2.
\label{dr7}
\end{equation}
On top of this, the thin slab and weak asymmetry approximations will be used. Let us introduce the notation

\begin{equation}
\varepsilon = \frac{x_0}{L}, \quad \mbox{and} \quad  v_{A2}^2=v_{A1}^2(1+\zeta),
\label{dr8}
\end{equation}
where $\zeta>0$ is taken, meaning $v_{A2}>v_{A1}$. The quantity $\varepsilon$ represents the ratio of the slab's half-width to its height, and $\zeta$ represents the asymmetry in the external Alfv\'en speeds. This asymmetry can be obtained through asymmetry in density, magnetic field strength, or both.

\subsubsection{Weak Asymmetry Approximation}

The asymmetry included in the model is crucial to enable the application of new solar magneto-seismology techniques. In particular, by introducing weak asymmetry, it will now be possible to make analytical progress in deriving the eigenfrequencies that the magnetic slab equilibrium supports, while also keeping the added complexity of having different external quantities on either side of the slab. This approach also allows for a comparison to the solutions presented in the symmetric model \citep{rob-82}.

The weak asymmetry assumption can be mathematically captured by $\zeta \ll 1$, which we then use to simplify the dispersion relation, Equation (\ref{dr6}), for analytical progress. The dispersion relation to leading-order in $\zeta$ is then

\begin{equation}
\left(k^2v_{A0}^2-\omega^2\right)\left(\frac{\rho_0m_1}{\rho_1(k^2v_{A1}^2-\omega^2)}+\frac{\rho_0m_2}{\rho_2(k^2v_{A2}^2-\omega^2)}\right) + 2m_0\left(\begin{matrix} \tanh{} \\ \coth{} \\ \end{matrix}\right) (m_0x_0) \approx 0,
\label{dr9}
\end{equation} 
where $\zeta$ is not explicitly written, however, it is within the quantities $\rho_2$, $m_2$ and $v_{A2}$ implicitly.

Equation (\ref{dr9}) is analogous to Equation (20) in \cite{zsam-all-18} (with the condition that $c_1=c_2$ is taken here), where now it is assumed in addition that the external densities, pressures and magnetic fields are of the same order. Let us now elaborate on this by the implementation of the small quantity $\zeta$. If we, again, compare Equation (\ref{dr9}) to the corresponding dispersion relation for a magnetic slab embedded in a symmetric environment, we note that the '$\tanh{}$' and '$\coth{}$' terms would correspond to sausage and kink modes, respectively \citep{rob-82}. To be consistent with notion, the terminology used in both \cite{all-erd-17} and \cite{zsam-all-18} will now be employed here. More precisely, in Equation (\ref{dr9}) the '$\tanh{}$' term corresponds to quasi-sausage and the '$\coth{}$' term to quasi-kink modes.

\subsubsection{Thin Slab Approximation}
The thin slab approximation is expressed by $\varepsilon \ll 1$, and we can use this condition to simplify the dispersion relation Equation (\ref{dr9}) further. The dispersion relation for quasi-sausage modes becomes

\begin{equation}
\left(k^2v_{A0}^2-\omega^2\right)\left(\frac{\rho_0m_1}{\rho_1(k^2v_{A1}^2-\omega^2)}+\frac{\rho_0m_2}{\rho_2(k^2v_{A2}^2-\omega^2)}\right) + 2m_0^2x_0\left(1-\frac{m_0^2x_0^2}{3}\right) \approx 0,
\label{dr10}
\end{equation} 
and for the quasi-kink oscillations it is

\begin{equation}
\left(k^2v_{A0}^2-\omega^2\right)\left(\frac{\rho_0m_1}{\rho_1(k^2v_{A1}^2-\omega^2)}+\frac{\rho_0m_2}{\rho_2(k^2v_{A2}^2-\omega^2)}\right) + \frac{2}{x_0} \approx 0.
\label{dr11}
\end{equation}

\subsubsection{Ordering of $\varepsilon$ and $\zeta$}
In the following, $O(\zeta) \sim O(\varepsilon)$ will be taken in order to use Equations (\ref{dr10}) and (\ref{dr11}) to derive the frequencies and keep second-order terms of size $\varepsilon \zeta$ and $\varepsilon^2$. We have ignored terms of size $\zeta^2$ in order to obtain the decoupled dispersion relations for the quasi-sausage and -kink modes. These assumptions are made in order to include the leading-order terms that are related to asymmetry.

\section{Eigenfrequencies of Standing Harmonic Modes} \label{Frequencies of the Standing Harmonic Modes}

The focus in this section is to derive the eigenfrequencies of the asymmetric magnetised slab system analytically. The eigenfrequency is a popular measurable quantity, to determine in the observations, and we study it in order to identify how sensitive the eigenfrequencies are to changes in the external environment. We will first derive the frequencies of the standing modes that may exist as functions of the small quantities $\varepsilon$ and $\zeta$, using a similar approach to that of \cite{ox-20}. The precise details of the method are not included here, and more information can be found in \cite{ox-20}. Care must be taken when using the solutions, as it was assumed that $m_0^2,m_1^2,m_2^2>0$, so the solutions are only valid when the characteristic speeds are ordered such that these relations are satisfied. In the case of the quasi-sausage waves, this will be followed by determining the frequency ratio of the first harmonic to the fundamental mode. The frequency ratio will not be calculated for the quasi-kink frequencies, with the reasons discussed in Section \ref{QK1}. We will use the quantisation of $k$ given in Equation (\ref{dr4}), and express the frequencies of the standing harmonic modes as $\omega_n$.

\subsection{Quasi-sausage Modes} \label{QS1}

The most simple solution to Equation (\ref{dr10}) is $\omega=kv_{A0}$, but we will ignore this case. This solution would mean $m_0^2=0$, and using this in Equation (\ref{dr5}) along with the requirement that the solution is evanescent, would give a trivial solution.

Introduce the notation

\begin{equation}
\bar{\Pi}=-\frac{\rho_1c_0^2(c_{T0}^2-v_{A1}^2)^{1/2}(c_{T0}^2-c_{T1}^2)^{1/2}(c_1^2+v_{A1}^2)^{1/2}}{\rho_0v_{A0}^2(c_1^2-c_{T0}^2)^{1/2}(c_0^2+v_{A0}^2)}.
\label{fi1}
\end{equation}
Then, by examining Equation (\ref{dr10}), the simplified dispersion relation for quasi-sausage waves, we find the following solution:

\begin{multline}
\omega_n^2 \approx \frac{n^2\pi^2c_{T0}^2}{L^2}\Bigg\{1+\varepsilon n\pi \bar{\Pi} \\ + \varepsilon^2n^2\pi^2 \left[\bar{\Pi}^2 c_{T0}^2\left(\frac{1}{2(c_1^2-c_{T0}^2)}-\frac{1}{2(c_{T1}^2-c_{T0}^2)}-\frac{1}{2(v_{A1}^2-c_{T0}^2)}-\frac{1}{(c_0^2-c_{T0}^2)}\right)+\frac{c_{T0}^2}{3(c_0^2+v_{A0}^2)}\right]\\ + \varepsilon\zeta\frac{n\pi \bar{\Pi}}{4}\left[\frac{v_{A1}^2}{v_{A1}^2-c_{T0}^2}\left(1+\frac{(c_1^2-c_{T0}^2)(v_{A1}^2-c_{T0}^2)}{(c_{T1}^2-c_{T0}^2)(c_1^2+v_{A1}^2)}\right)-\frac{2v_{A1}^2\gamma}{2c_1^2+v_{A1}^2\gamma}\right]\Bigg\}, \\
\label{fi2}
\end{multline}
valid when $v_{A1}<c_{T0}<c_1$, to ensure that $m_i^2>0$ for $i=0,1,2$. Note the form in which we present the solutions suggests the ordering $c_1<c_{T0}<v_{A1}$ could be valid. However, the quantity $\bar{\Pi}$ has been simplified using the ordering $v_{A1}<c_{T0}<c_1$, and by deriving the more general form of this quantity, it is observed that the ordering $c_1<c_{T0}<v_{A1}$ would result in a solution for the frequency where the condition $m_i^2>0$ for $i=0,1,2$ is not satisfied. We therefore do not include this latter solution in our study as we only consider surface waves.

The derivation of eigenfrequencies is analogous to that in the symmetric slab model presented in \cite{rob-82}. In that work, multiple solutions are presented, as the frequency is different depending on the orderings of the characteristic speeds. The solution given by Equation (\ref{fi2}) is analogous to Equation (16d) in \cite{rob-82}. While it is possible to derive analogues of Equations (16b, c) under speed orderings similar to the ones presented there, those solutions describe a surface wave, which, in a thin slab, changes its character to that of a body wave. Excluding these types of modes, we now proceed only with the analysis of solutions that have the character of surface waves in both wide and thin slabs, and therefore satisfy $m_i^2>0$ for $i=0,1,2$.

A comparison to the magnetic slab embedded in a non-magnetic asymmetric environment, studied by \cite{ox-20}, can also be made here. If we were to set $B_1=B_2=0$ and consequently $v_{A1}=v_{A2}=0$, then the quantity $\bar{\Pi}$ would reduce to the quantity $-\Pi$, where $\Pi$ is given by \cite{ox-20}. It can then be shown that the terms of order $1$, $\varepsilon$ and $\varepsilon^2$ in Equation (\ref{fi2}) are each equal to the corresponding terms of the same order in Equation (26) in \cite{ox-20}. Although the term of order $\varepsilon \zeta$ here shares a similar form to the term of order $\varepsilon \delta$ in \cite{ox-20}, one does not reduce to the other as $\zeta$ and $\delta$ represent different types of asymmetry.

Let us now express the frequency ratio, a popular quantity in SMS studies:

\begin{multline}
\frac{\omega_2}{\omega_1} \approx 2\Bigg\{1 + \varepsilon\frac{\pi \bar{\Pi}}{2} + \varepsilon^2\pi^2\left[\frac{\bm{\bar{\Pi}^2}}{8}\left(\frac{6c_{T0}^2}{c_1^2-c_{T0}^2}-\frac{6c_{T0}^2}{c_{T1}^2-c_{T0}^2}-\frac{6c_{T0}^2}{v_{A1}^2-c_{T0}^2}-\frac{12c_{T0}^2}{c_0^2-c_{T0}^2}-5\right)+\frac{c_{T0}^2}{2(c_0^2+v_{A0}^2)}\right] \\ + \varepsilon\zeta\frac{\pi \bar{\Pi}}{8}\left[\frac{v_{A1}^2}{v_{A1}^2-c_{T0}^2}\left(1+\frac{(c_1^2-c_{T0}^2)(v_{A1}^2-c_{T0}^2)}{(c_{T1}^2-c_{T0}^2)(c_1^2+v_{A1}^2)}\right)-\frac{2v_{A1}^2\gamma}{2c_1^2+v_{A1}^2\gamma}\right]\Bigg\}. \\
\label{fi3}
\end{multline}
The frequency ratio given by Equation (\ref{fi3}), here, can be compared with the frequency ratio given by Equation (27) in \cite{ox-20} in exactly the same way as the frequencies have been compared in the above discussion.

Next, there is another solution to Equation (\ref{dr10}), but let us first define $\tilde{\Pi}$ by

\begin{equation}
\tilde{\Pi}=\frac{\rho_1^2c_1^2(c_0^2-c_1^2)^2(v_{A1}^2-c_1^2)}{\rho_0^2(c_{T0}^2-c_1^2)^2(c_0^2+v_{A0}^2)^2}.
\label{fi5}
\end{equation}
Then, the solution to Equation (\ref{dr10}) up to and including the first term involving $\zeta$ characterising the external asymmetry is

\begin{equation}
\omega_n^2 \approx \frac{n^2\pi^2c_1^2}{L^2}\Bigg\{1+\varepsilon^2 n^2\pi^2 \tilde{\Pi} + \varepsilon^2\zeta\frac{ n^2\pi^2 \tilde{\Pi}v_{A1}^2}{2}\left[\frac{1}{v_{A1}^2-c_1^2}-\frac{2\gamma}{2c_1^2+\gamma v_{A1}^2}\right]\Bigg\}.
\label{fi6}
\end{equation}
This solution is analogous to that obtained by \cite{rob-82} in their Equation (16a) for the symmetric slab model, and one can reduce to that solution by taking $\zeta=0$ here. For this frequency, a comparison to \cite{ox-20} cannot be made. This frequency can only exist due to the condition that the external plasma is isothermal, which is not the case in the study of \cite{ox-20}.

The frequency ratio is then

\begin{equation}
\frac{\omega_2}{\omega_1} \approx 2\left\{1+\varepsilon^2\frac{3\pi^2\tilde{\Pi}}{2}+ \varepsilon^2\zeta\frac{ 3\pi^2 \tilde{\Pi}v_{A1}^2}{4}\left[\frac{1}{v_{A1}^2-c_1^2}-\frac{2\gamma}{2c_1^2+\gamma v_{A1}^2}\right]\right\}.
\label{fi7}
\end{equation}

Illustrations of the frequency ratio of the first harmonic to the fundamental mode of the asymmetric magnetic slab equilibrium system are given in Figures \ref{fig:ratio_epsilon} and \ref{fig:ratio_zeta}. For both solutions, we see a quadratic-like dependency on $\varepsilon$ as expected, whereas a linear relationship with $\zeta$ is observed, which is weak. This is because the first $\zeta$-term in the solution given by Equation (\ref{fi3}) appears as a second-order quantity in the small parameters $\zeta$ and $\varepsilon$, and the first $\zeta$-term in the solution given by Equation (\ref{fi7}) appears as a third-order quantity in the small parameters $\zeta$ and $\varepsilon$. When comparing how changes in the magnetic asymmetry and the slab width affect the eigenfrequency ratio, it is evident that changes to the magnetic asymmetry will have a less significant effect. The values for the background parameters of the system have been chosen to satisfy the ordering necessary for $m_0^2>0$ and $m_1^2>0$ and to demonstrate the dependency of both solutions on $\varepsilon$ and $\zeta$ clearly.

In addition to the analytical solutions, some numerical solutions have been included in the plots of Equations (\ref{fi3}) and (\ref{fi7}). These numerical solutions are included to indicate what values of $\varepsilon$ and $\zeta$ may be considered small enough to use the analytical solutions. For example, a reasonable 'cut-off' for the parameter $\varepsilon$ may be when the analytical solution has deviated from the numerical solution by $20\%$. This first occurs around $\varepsilon=0.2$, and as $\varepsilon$ is increased past this value, the analytical solutions deviate significantly from the numerical solutions. For Figure \ref{fig:ratio_epsilon}, the parameter $\varepsilon$ ranges from $0$ to $0.3$. However, the first small portion of the parameter range must be discarded (e.g. $\varepsilon < 0.02 $) as for this plot $\zeta=0.1$ and the ordering $\zeta^2 \ll \varepsilon$ must be obeyed for the solutions to be valid. A similar consideration must also be made for Figure \ref{fig:ratio_zeta}.

Illustrations of the fundamental and first harmonic standing quasi-sausage waves are given in Figures \ref{fig:standing_saus_fund} and \ref{fig:standing_saus_first}.

\begin{figure}
\hspace{0.2cm}
  \begin{minipage}[t]{0.45\textwidth}
\centering
\scalebox{0.7}[0.7]{
\begin{tikzpicture}
%middle
\path [fill=lightgray, opacity=0.6] (3.5,5) -- (3.5,6) -- (6.5,6) -- (6.5,5) -- (3.5,5);
\path [fill=lightgray, opacity=0.6] (3.5,0) -- (3.5,1) -- (6.5,1) -- (6.5,0) -- (3.5,0);
\path [fill=lightgray, opacity=0.6] (6.5,1) to [out=65, in=-65] (6.5,5) to (3.5,5) to [out=-135, in=135]  (3.5,1) to (6.5,1);

%curved surfaces middle
\shade[left color=darkgray,right color=white, opacity=0.99] (6.5,0) -- (6.5,1) -- (7,1.3) -- (7,0.3) -- (6.5,0);
\shade[left color=darkgray,right color=white, opacity=0.99] (6.5,5) -- (6.5,6) -- (7,6.3) -- (7,5.3) -- (6.5,5);
%\shade[left color=darkgray,right color=white, opacity=0.3] (3.5,1) to [out=135, in=-90] (2.75,2.125) to [out=90, in=-135] (3.5,3) to [out=45, in=-90] (4.25, 3.875) to [out=90, in=-45] (3.5,5) to (4,5.3) to [out=-45, in=90] (4.75,4.175) to [out=-90, in=45] (4,3.3) to [out=-135, in=90] (3.25,2.425) to [out=-90, in=135] (4,1.3) to (3.5,1);
\shade[left color=darkgray,right color=white, opacity=0.3] (3.5,1) to [out=142, in=-142] (3.5,5) to (4,5.3) to [out=-140, in=140]  (4,1.3) to (3.5,1);
%\shade[top color=white,bottom color=lightgray, opacity=0.7] (-0.5,1) -- (0.,1.3) -- (4,1.3) -- (3.5,1) -- (-0.5,1);
\shade[bottom color=lightgray,top color=white, opacity=0.45] (3.5,1) -- (6.5,1) -- (6.5,1.3) -- (4.,1.3) -- (3.5,1);
\shade[left color=darkgray,right color=white, opacity=0.99] (6.5,1) to [out=65, in=-65]  (6.5,5) to (7,5.3) to [out=-65, in=65]  (7,1.3) to (6.5,1);

%middle top and bottom
\shade[top color=lightgray,bottom color=white, opacity=0.45] (3.5,0) -- (6.5,0) -- (6.5,-0.3) -- (3.5,-0.3) -- (3.5,0);
\shade[top color=white,bottom color=lightgray, opacity=0.45] (3.5,6) -- (4,6.3) -- (7,6.3) -- (6.5,6) -- (3.5,6);
\shade[top color=white,bottom color=lightgray, opacity=0.45] (3.5,5) -- (4,5.3) -- (7,5.3) -- (6.5,5) -- (3.5,5);
\shade[left color=lightgray,right color=white, opacity=0.99] (6.5,0) -- (6.5,-0.3) -- (7,0) -- (7,0.3) -- (6.5,0);

%curved surface left + very top + very bottom right
\shade[top color=lightgray,bottom color=white, opacity=0.8] (6.5,0) -- (10,0) -- (10,-0.3) -- (6.5,-0.3) -- (6.5,0);
\shade[top color=white,bottom color=lightgray, opacity=0.8] (6.5,6) -- (7,6.3) -- (10.5,6.3) -- (10,6) -- (6.5,6);
\shade[right color=white,left color=lightgray, opacity=0.8] (10,-0.) -- (10,1) -- (10.5,1.3) -- (10.5,0.3) -- (10,-0.);
\shade[bottom color=white,top color=lightgray, opacity=0.3] (10,-0.) -- (10,1) -- (10.5,1.3) -- (10.5,0.3) -- (10,-0.);
\shade[right color=white, left color=lightgray, opacity=0.8] (10,5.) -- (10,6) -- (10.5,6.3) -- (10.5,5.3) -- (10,5.);
\shade[left color=lightgray,right color=white, opacity=0.5] (-0.5,1) to (-1,2) to (-0.5,3.) to (-1,4) to (-0.5,5) to (0,5.3) to (-0.5,4.3) to (0,3.3) to (-0.5,2.3) to (0,1.3);
\shade[top color=lightgray,bottom color=white, opacity=0.3] (10,-0.3) -- (10,0.) -- (10.5,0.3) -- (10.5,-0.) -- (10,-0.3);
\shade[left color=lightgray, right color=white, opacity=0.3] (10,-0.3) -- (10,0.) -- (10.5,0.3) -- (10.5,-0.) -- (10,-0.3);

%curved surface right
\shade[top color=lightgray,bottom color=white, opacity=0.3] (-0.5,0) -- (3.5,0) -- (3.5,-0.3) -- (-0.5,-0.3) -- (-0.5,0);
\shade[top color=white,bottom color=lightgray, opacity=0.3] (-0.5,6) -- (0.,6.3) -- (4,6.3) -- (3.5,6) -- (-0.5,6);
\shade[top color=white,bottom color=lightgray, opacity=0.7] (-0.5,5) -- (0.,5.3) -- (4,5.3) -- (3.5,5) -- (-0.5,5);
\shade[right color=white,left color=lightgray, opacity=0.8] (10,1) to (9.5,2) to (10,3) to (9.5,4) to (10,5) to (10.5,5.3) to (10,4.3) to (10.5,3.3) to (10,2.3) to (10.5,1.3) to (10,1);

%top and bottom corners right
\shade[top color=white,bottom color=black, opacity=0.9] (6.5,1) -- (7.,1.3) -- (10.5,1.3) -- (10,1) -- (6.5,1);
\shade[top color=white,bottom color=black, opacity=0.9] (6.5,5) -- (7.,5.3) -- (10.5,5.3) -- (10,5) -- (6.5,5);

%top dashed left
\draw [ultra thick, dashed] (-0.5,5.) -- (3.5,5);
\draw [ultra thick, dashed, path fading = east] (-0.5,5.) -- (0,5.3);
\draw [ultra thick, dashed, path fading = east] (0.5,5.) -- (1,5.3);
\draw [ultra thick, dashed, path fading = east] (1.5,5.) -- (2,5.3);
\draw [ultra thick, dashed, path fading = east] (2.5,5.) -- (3,5.3);

%bottom dashed left
\draw [ultra thick, dashed] (-0.5,1.) -- (3.5,1);
\draw [ultra thick, dashed, path fading = east] (-0.5,1.) -- (0,1.3);
\draw [ultra thick, dashed, path fading = east] (0.5,1.) -- (1,1.3);
\draw [ultra thick, dashed, path fading = east] (1.5,1.) -- (2,1.3);
\draw [ultra thick, dashed, path fading = east] (2.5,1.) -- (3,1.3);

%top and bottom dashed middle
\draw [ultra thick, dashed] (3.5,5) --(6.5,5);
\draw [ultra thick, dashed, path fading=east] (4.5,5) -- (5,5.3);
\draw [ultra thick, dashed, path fading=east] (5.5,5) -- (6,5.3);
\draw [ultra thick, dashed] (3.5,1) --(6.5,1);
\draw [ultra thick, dashed, path fading=east] (4.5,1) -- (5,1.3);
\draw [ultra thick, dashed, path fading=east] (5.5,1) -- (6,1.3);

%1 filling
\shade[top color=lightgray, bottom color=lightgray, opacity=0.9] (6.5,1.) to (6.5,0.) to (10,0) to (10,1) to (6.5,1);
\shade[top color=lightgray, bottom color=lightgray, opacity=0.9] (6.5,1) to [out=65,in=-65] (6.5,5) to (10,5) to (9.5,4) to (10,3) to (9.5,2) to (10,1) to (6.5,1);
%\shade[top color=lightgray, bottom color=lightgray, opacity=0.9] (6.5,5.) to (10,5.) to (9.5,4) to (10,3) to (6.5,3) to [out=135,in=-135] (6.5,5);
\shade[top color=lightgray, bottom color=lightgray, opacity=0.9] (6.5,5.) to (6.5,6.) to (10,6) to (10,5) to (6.5,5);

%2 filling
\shade[top color=lightgray, bottom color=lightgray, opacity=0.3] (-0.5,1.) to (-0.5,0.) to (3.5,0) to (3.5,1) to (-0.5,1);
\shade[top color=lightgray, bottom color=lightgray, opacity=0.3] (-0.5,1) to (-1,2) to (-0.5,3) to (-1,4) to (-0.5,5)  to (3.5,5) to [out=-140,in=140] (3.5,1) to (-0.5,1);
%\shade[top color=lightgray, bottom color=lightgray, opacity=0.3] (-0.5,5.) to (3.5,5.) to [out=-45,in=90]
 (4.25,3.875) to [out=-90, in=45] (3.5,3) to (-0.5,3) to (-1,4) to (-0.5,5);
\shade[top color=lightgray, bottom color=lightgray, opacity=0.3] (-0.5,5.) to (-0.5,6.) to (3.5,6) to (3.5,5) to (-0.5,5);

% magnetic field arrows middle
\draw [ultra thick, red, -stealth] (6.16,4)  to  (6.14,4.1);
\draw [ultra thick, red] (6.25,3)  to [out=90,in=-70]  (5.85,5);
\draw [ultra thick, red] (5.85,1)  to [out=70,in=-90] (6.25,3);
\draw [ultra thick, red, -stealth] (6.1,1.8)  to  (6.12,1.9);

\draw [ultra thick, red, -stealth] (5.1,4)  to  (5.1,4.1);
\draw [ultra thick, red, path fading = south] (5.1,3.25)  to  (5.1,5);
\draw [ultra thick, red, path fading = north] (5.1,1)  to (5.1,2.75);
\draw [ultra thick, red, -stealth] (5.1,1.8)  to  (5.1,1.9);

\draw [ultra thick, red, -stealth] (3.73,4)  to  (3.76,4.1);
\draw [ultra thick, red] (3.55,3)  to [out=90,in=-120]  (4.25,5);
\draw [ultra thick, red] (4.25,1)  to [out=120,in=-90] (3.55,3);
\draw [ultra thick, red, -stealth] (3.81,1.8)  to  (3.76,1.9);

%magnetic field arrows left
\draw [ultra thick, red, -stealth] (0.48,4)  to  (0.5,4.1);
\draw [ultra thick, red, path fading = south] (0.4,3.25)  to [out=90,in=-115]  (0.85,5);
\draw [ultra thick, red, path fading = north] (0.85,1)  to [out=115,in=-90] (0.4,2.75);
\draw [ultra thick, red, -stealth] (0.54,1.8)  to  (0.5,1.9);

\draw [ultra thick, red, -stealth] (1.93,4)  to  (1.96,4.1);
\draw [ultra thick, red, path fading = south] (1.75,3.25)  to [out=85,in=-120]  (2.45,5);
\draw [ultra thick, red, path fading = north] (2.45,1)  to [out=120,in=-85] (1.75,2.75);
\draw [ultra thick, red, -stealth] (2.01,1.8)  to  (1.96,1.9);

%magnetic field arrows right
\draw [ultra thick, red, -stealth] (7.56,4)  to  (7.54,4.1);
\draw [ultra thick, red, path fading = south] (7.65,3.25)  to [out=90,in=-70]  (7.25,5);
\draw [ultra thick, red, path fading = north] (7.25,1)  to [out=70,in=-90] (7.65,2.75);
\draw [ultra thick, red, -stealth] (7.5,1.8)  to  (7.52,1.9);

\draw [ultra thick, red, -stealth] (8.02,4)  to  (8,4.1);
\draw [ultra thick, red, path fading = south] (8.1,3.25)  to [out=90,in=-72.5]  (7.75,5);
\draw [ultra thick, red, path fading = north] (7.75,1)  to [out=72.5,in=-90] (8.1,2.75);
\draw [ultra thick, red, -stealth] (7.97,1.8)  to  (7.99,1.9);

\draw [ultra thick, red, -stealth] (8.48,4)  to  (8.46,4.1);
\draw [ultra thick, red, path fading = south] (8.55,3.25)  to [out=90,in=-75]  (8.25,5);
\draw [ultra thick, red, path fading = north] (8.25,1)  to [out=75,in=-90] (8.55,2.75);
\draw [ultra thick, red, -stealth] (8.44,1.8)  to  (8.46,1.9);

\draw [ultra thick, red, -stealth] (8.94,4)  to  (8.92,4.1);
\draw [ultra thick, red, path fading = south] (9,3.25)  to [out=90,in=-77.5]  (8.75,5);
\draw [ultra thick, red, path fading = north] (8.75,1)  to [out=77.5,in=-90] (9,2.75);
\draw [ultra thick, red, -stealth] (8.92,1.8)  to  (8.94,1.9);

\draw [ultra thick, red, -stealth] (9.41,4)  to  (9.4,4.1);
\draw [ultra thick, red, path fading = south] (9.45,3.25)  to [out=90,in=-80]  (9.25,5);
\draw [ultra thick, red, path fading = north] (9.25,1)  to [out=80,in=-90] (9.45,2.75);
\draw [ultra thick, red, -stealth] (9.39,1.8)  to  (9.4,1.9);

%curved dashed left vertical
\draw [ultra thick, dashed] (3.5,5) --(3.5,6);
\draw [ultra thick, dashed] (3.5,1) to [out=140,in=-140] (3.5,5);
%\draw [ultra thick, dashed] (3.5,1) to [out=135,in=-90]  (2.75,2.125) to [out=90,in=-135]  (3.5,3);
\draw [ultra thick, dashed] (3.5,0) --(3.5,1);

%curved dashed left horizontal
\draw [ultra thick, dashed, path fading=east] (3.5,6) -- (4,6.3);
\draw [ultra thick, dashed, path fading=east] (3.5,5) -- (4,5.3);
\draw [ultra thick, dashed, path fading=east] (2.8,4) -- (3.3,4.3);
\draw [ultra thick, dashed, path fading=east] (2.6,3) -- (3.1,3.3);
\draw [ultra thick, dashed, path fading=east] (2.75,2) -- (3.25,2.3);
\draw [ultra thick, dashed, path fading=east] (3.5,1) -- (4,1.3);
\draw [ultra thick, dashed, path fading=east] (3.5,0) -- (4,0.3);

%bottom and top dashed right
\draw [ultra thick, dashed] (6.5,5) --(10,5);
\draw [ultra thick, dashed, path fading=east] (7.5,5) -- (8,5.3);
\draw [ultra thick, dashed, path fading=east] (8.5,5) -- (9,5.3);
\draw [ultra thick, dashed, path fading=east] (9.5,5) -- (10,5.3);
\draw [ultra thick, dashed] (6.5,1) --(10,1);
\draw [ultra thick, dashed, path fading=east] (7.5,1) -- (8,1.3);
\draw [ultra thick, dashed, path fading=east] (8.5,1) -- (9,1.3);
\draw [ultra thick, dashed, path fading=east] (9.5,1) -- (10,1.3);

%curved dashed right
\draw [ultra thick, dashed] (6.5,5.) -- (6.5,6);
\draw [ultra thick, dashed] (6.5,1)  to [out=65,in=-65] (6.5,5);
%\draw [ultra thick, dashed] (6.5,1)  to [out=45,in=-45] (6.5,3);
\draw [ultra thick, dashed] (6.5,0.) -- (6.5,1);
\draw [ultra thick, dashed, path fading=east] (6.5,0) -- (7,0.3);
\draw [ultra thick, dashed, path fading=east] (6.5,6) -- (7,6.3);
\draw [ultra thick, dashed, path fading=east] (6.5,5) -- (7,5.3);
\draw [ultra thick, dashed, path fading=east] (6.9,4) -- (7.4,4.3);
\draw [ultra thick, dashed, path fading=east] (7,3) -- (7.5,3.3);
\draw [ultra thick, dashed, path fading=east] (7,2) -- (7.5,2.3);
\draw [ultra thick, dashed, path fading=east] (6.5,1) -- (7,1.3);

\draw [<-] (0,6) -- (0,-0.3); 	%z
\draw [->] (0,1) -- (10,1);		%x
\draw [->] (0,1) -- (0.5,1.3);	%y

\small

\node [below] at (10,1) {$x$};
\node [left] at (0,6) {$z$};
\node [right] at (0.45,1.3) {$y$};

\large
\node [right] at (0.2,3) {$\rho_1$,$p_1$,$T_1$,$B_1$};
\node [right] at (3.75,3) {$\rho_0$,$p_0$,$T_0$,$B_0$};
\node [right] at (7.4,3) {$\rho_2$,$p_2$,$T_2$,$B_2$};
\node [left] at (-0.5,1) {$z=0$};
\node [left] at (-0.5,5) {$z=L$};
\node [below] at (3.5,-0.2) {$x=-x_0$};
\node [below] at (6.5,-0.2) {$x=x_0$};

%magnetic field end points middle
\fill[red] (4.25,5) circle (0.08cm);
\fill[red] (4.25,1) circle (0.08cm);
\fill[red] (5.1,5) circle (0.08cm);
\fill[red] (5.1,1) circle (0.08cm);
\fill[red] (5.85,5) circle (0.08cm);
\fill[red] (5.85,1) circle (0.08cm);

%magnetic field end points left
\fill[red] (0.85,5) circle (0.08cm);
\fill[red] (0.85,1) circle (0.08cm);
\fill[red] (2.45,5) circle (0.08cm);
\fill[red] (2.45,1) circle (0.08cm);

%magnetic field end points right
\fill[red] (7.25,5) circle (0.08cm);
\fill[red] (7.25,1) circle (0.08cm);
\fill[red] (7.75,5) circle (0.08cm);
\fill[red] (7.75,1) circle (0.08cm);
\fill[red] (8.25,5) circle (0.08cm);
\fill[red] (8.25,1) circle (0.08cm);
\fill[red] (8.75,5) circle (0.08cm);
\fill[red] (8.75,1) circle (0.08cm);
\fill[red] (9.25,5) circle (0.08cm);
\fill[red] (9.25,1) circle (0.08cm);
\end{tikzpicture}
}
\caption{Illustration of a fundamental standing quasi-sausage mode oscillation in the magnetic slab embedded in a magnetically asymmetric environment.}
\label{fig:standing_saus_fund}
  \end{minipage}
 \hspace{0.5cm}
  \begin{minipage}[t]{0.45\textwidth}
\centering
\scalebox{0.7}[0.7]{
\begin{tikzpicture}
%middle
\path [fill=lightgray, opacity=0.6] (3.5,5) -- (3.5,6) -- (6.5,6) -- (6.5,5) -- (3.5,5);
\path [fill=lightgray, opacity=0.6] (3.5,0) -- (3.5,1) -- (6.5,1) -- (6.5,0) -- (3.5,0);
\path [fill=lightgray, opacity=0.6] (6.5,1) to [out=45, in=-45] (6.5, 3.)  to [out=135, in=-135] (6.5,5) to (3.5,5) to [out=-45, in=90] (4.25, 3.875) to [out=-90, in=45] (3.5,3) to [out=-135, in=90] (2.75,2.125) to [out=-90, in=135] (3.5,1) to (6.5,1);

%curved surfaces middle
\shade[left color=darkgray,right color=white, opacity=0.99] (6.5,0) -- (6.5,1) -- (7,1.3) -- (7,0.3) -- (6.5,0);
\shade[left color=darkgray,right color=white, opacity=0.99] (6.5,5) -- (6.5,6) -- (7,6.3) -- (7,5.3) -- (6.5,5);
\shade[left color=darkgray,right color=white, opacity=0.3] (3.5,1) to [out=135, in=-90] (2.75,2.125) to [out=90, in=-135] (3.5,3) to [out=45, in=-90] (4.25, 3.875) to [out=90, in=-45] (3.5,5) to (4,5.3) to [out=-45, in=90] (4.75,4.175) to [out=-90, in=45] (4,3.3) to [out=-135, in=90] (3.25,2.425) to [out=-90, in=135] (4,1.3) to (3.5,1);
%\shade[top color=white,bottom color=lightgray, opacity=0.7] (-0.5,1) -- (0.,1.3) -- (4,1.3) -- (3.5,1) -- (-0.5,1);
\shade[bottom color=lightgray,top color=white, opacity=0.45] (3.5,1) -- (6.5,1) -- (6.5,1.3) -- (4.,1.3) -- (3.5,1);
\shade[left color=darkgray,right color=white, opacity=0.99] (6.5,1) to [out=45, in=-45] (6.5,3) to [out=135, in=-135] (6.5,5) to (7,5.3) to [out=-135, in=135] (7,3.3) to [out=-45, in=45] (7,1.3) to (6.5,1);

%middle top and bottom
\shade[top color=lightgray,bottom color=white, opacity=0.45] (3.5,0) -- (6.5,0) -- (6.5,-0.3) -- (3.5,-0.3) -- (3.5,0);
\shade[top color=white,bottom color=lightgray, opacity=0.45] (3.5,6) -- (4,6.3) -- (7,6.3) -- (6.5,6) -- (3.5,6);
\shade[top color=white,bottom color=lightgray, opacity=0.45] (3.5,5) -- (4,5.3) -- (7,5.3) -- (6.5,5) -- (3.5,5);
\shade[left color=lightgray,right color=white, opacity=0.99] (6.5,0) -- (6.5,-0.3) -- (7,0) -- (7,0.3) -- (6.5,0);

%curved surface left + very top + very bottom right
\shade[top color=lightgray,bottom color=white, opacity=0.8] (6.5,0) -- (10,0) -- (10,-0.3) -- (6.5,-0.3) -- (6.5,0);
\shade[top color=white,bottom color=lightgray, opacity=0.8] (6.5,6) -- (7,6.3) -- (10.5,6.3) -- (10,6) -- (6.5,6);
\shade[right color=white,left color=lightgray, opacity=0.8] (10,-0.) -- (10,1) -- (10.5,1.3) -- (10.5,0.3) -- (10,-0.);
\shade[bottom color=white,top color=lightgray, opacity=0.3] (10,-0.) -- (10,1) -- (10.5,1.3) -- (10.5,0.3) -- (10,-0.);
\shade[right color=white, left color=lightgray, opacity=0.8] (10,5.) -- (10,6) -- (10.5,6.3) -- (10.5,5.3) -- (10,5.);
\shade[left color=lightgray,right color=white, opacity=0.5] (-0.5,1) to (-1,2) to (-0.5,3.) to (-1,4) to (-0.5,5) to (0,5.3) to (-0.5,4.3) to (0,3.3) to (-0.5,2.3) to (0,1.3);
\shade[top color=lightgray,bottom color=white, opacity=0.3] (10,-0.3) -- (10,0.) -- (10.5,0.3) -- (10.5,-0.) -- (10,-0.3);
\shade[left color=lightgray, right color=white, opacity=0.3] (10,-0.3) -- (10,0.) -- (10.5,0.3) -- (10.5,-0.) -- (10,-0.3);

%curved surface right
\shade[top color=lightgray,bottom color=white, opacity=0.3] (-0.5,0) -- (3.5,0) -- (3.5,-0.3) -- (-0.5,-0.3) -- (-0.5,0);
\shade[top color=white,bottom color=lightgray, opacity=0.3] (-0.5,6) -- (0.,6.3) -- (4,6.3) -- (3.5,6) -- (-0.5,6);
\shade[top color=white,bottom color=lightgray, opacity=0.7] (-0.5,5) -- (0.,5.3) -- (4,5.3) -- (3.5,5) -- (-0.5,5);
\shade[right color=white,left color=lightgray, opacity=0.8] (10,1) to (9.5,2) to (10,3) to (9.5,4) to (10,5) to (10.5,5.3) to (10,4.3) to (10.5,3.3) to (10,2.3) to (10.5,1.3) to (10,1);

%top and bottom corners right
\shade[top color=white,bottom color=black, opacity=0.9] (6.5,1) -- (7.,1.3) -- (10.5,1.3) -- (10,1) -- (6.5,1);
\shade[top color=white,bottom color=black, opacity=0.9] (6.5,5) -- (7.,5.3) -- (10.5,5.3) -- (10,5) -- (6.5,5);

%top dashed left
\draw [ultra thick, dashed] (-0.5,5.) -- (3.5,5);
\draw [ultra thick, dashed, path fading = east] (-0.5,5.) -- (0,5.3);
\draw [ultra thick, dashed, path fading = east] (0.5,5.) -- (1,5.3);
\draw [ultra thick, dashed, path fading = east] (1.5,5.) -- (2,5.3);
\draw [ultra thick, dashed, path fading = east] (2.5,5.) -- (3,5.3);

%bottom dashed left
\draw [ultra thick, dashed] (-0.5,1.) -- (3.5,1);
\draw [ultra thick, dashed, path fading = east] (-0.5,1.) -- (0,1.3);
\draw [ultra thick, dashed, path fading = east] (0.5,1.) -- (1,1.3);
\draw [ultra thick, dashed, path fading = east] (1.5,1.) -- (2,1.3);
\draw [ultra thick, dashed, path fading = east] (2.5,1.) -- (3,1.3);

%top and bottom dashed middle
\draw [ultra thick, dashed] (3.5,5) --(6.5,5);
\draw [ultra thick, dashed, path fading=east] (4.5,5) -- (5,5.3);
\draw [ultra thick, dashed, path fading=east] (5.5,5) -- (6,5.3);
\draw [ultra thick, dashed] (3.5,1) --(6.5,1);
\draw [ultra thick, dashed, path fading=east] (4.5,1) -- (5,1.3);
\draw [ultra thick, dashed, path fading=east] (5.5,1) -- (6,1.3);

%1 filling
\shade[top color=lightgray, bottom color=lightgray, opacity=0.9] (6.5,1.) to (6.5,0.) to (10,0) to (10,1) to (6.5,1);
\shade[top color=lightgray, bottom color=lightgray, opacity=0.9] (6.5,1) to [out=45,in=-45] (6.5,3) to (10,3) to (9.5,2) to (10,1) to (6.5,1);
\shade[top color=lightgray, bottom color=lightgray, opacity=0.9] (6.5,5.) to (10,5.) to (9.5,4) to (10,3) to (6.5,3) to [out=135,in=-135] (6.5,5);
\shade[top color=lightgray, bottom color=lightgray, opacity=0.9] (6.5,5.) to (6.5,6.) to (10,6) to (10,5) to (6.5,5);

%2 filling
\shade[top color=lightgray, bottom color=lightgray, opacity=0.3] (-0.5,1.) to (-0.5,0.) to (3.5,0) to (3.5,1) to (-0.5,1);
\shade[top color=lightgray, bottom color=lightgray, opacity=0.3] (-0.5,1) to (-1,2) to (-0.5,3) to (3.5,3) to [out=-135,in=90] (2.75,2.125) to [out=-90,in=135] 
 (3.5,1) to (-0.5,1);
\shade[top color=lightgray, bottom color=lightgray, opacity=0.3] (-0.5,5.) to (3.5,5.) to [out=-45,in=90]
 (4.25,3.875) to [out=-90, in=45] (3.5,3) to (-0.5,3) to (-1,4) to (-0.5,5);
\shade[top color=lightgray, bottom color=lightgray, opacity=0.3] (-0.5,5.) to (-0.5,6.) to (3.5,6) to (3.5,5) to (-0.5,5);

% magnetic field arrows middle
\draw [ultra thick, red, -stealth] (5.2,4)  to  (5.2,4.1);
\draw [ultra thick, red, path fading = south] (5.8,3.25)  to [out=125,in=-125]  (5.85,5);
\draw [ultra thick, red, path fading = north] (5.85,1)  to [out=55,in=-55] (6.05,2.75);
\draw [ultra thick, red, -stealth] (6.23,1.8)  to  (6.24,1.9);

\draw [ultra thick, red, -stealth] (5.53,4)  to  (5.52,4.1);
\draw [ultra thick, red, path fading = south] (5.2,3.25)  to (5.2,5);
\draw [ultra thick, red, path fading = north] (5.2,1)  to  (5.2,2.75);
\draw [ultra thick, red, -stealth] (5.2,1.8)  to  (5.2,1.9);

\draw [ultra thick, red, -stealth] (4.8,4)  to  (4.79,4.1);
\draw [ultra thick, red, path fading = south] (4.5,3.45)  to [out=30,in=-30]  (4.25,5);
\draw [ultra thick, red, path fading = north] (4.25,1)  to [out=135,in=-135] (3.55,2.55);
\draw [ultra thick, red, -stealth] (3.53,1.8)  to  (3.49,1.9);

%magnetic field arrows left
\draw [ultra thick, red, -stealth] (2.93,4)  to  (2.91,4.1);
\draw [ultra thick, red, path fading = south] (2.7,3.4)  to [out=34,in=-30]  (2.25,5);
%\draw [ultra thick, red] (1.75,2.75) to (2.7,3.4);
\draw [ultra thick, red, path fading = north] (2.25,1)  to [out=135,in=-146] (1.75,2.75);
\draw [ultra thick, red, -stealth] (1.65,1.8)  to  (1.59,1.9);

\draw [ultra thick, red, -stealth] (0.69,4)  to  (0.685,4.1);
\draw [ultra thick, red] (0.55,3.4)  to [out=65,in=-55]  (0.35,5);
\draw [ultra thick, red] (0.25,2.75) to (0.55,3.4);
\draw [ultra thick, red] (0.35,1)  to [out=115,in=-115] (0.25,2.75);
\draw [ultra thick, red, -stealth] (0.1,1.8)  to  (0.08,1.9);

% magnetic field arrows right
\draw [ultra thick, red, -stealth] (6.86,4)  to  (6.85,4.1);
\draw [ultra thick, red,  path fading = south] (7.1,3.25)  to [out=117,in=-125]  (7.15,5);
%\draw [ultra thick, red] (7.35,2.75) to (7.1,3.25);
\draw [ultra thick, red,  path fading = north] (7.15,1)  to [out=55,in=-63] (7.35,2.75);
\draw [ultra thick, red, -stealth] (7.5,1.8)  to  (7.51,1.9);

\draw [ultra thick, red, -stealth] (7.36,4)  to  (7.35,4.1);
\draw [ultra thick, red, path fading = south] (7.6,3.25)  to [out=120,in=-120]  (7.65,5);
%\draw [ultra thick, red] (1.75,2.75) to (2.7,3.4);
\draw [ultra thick, red, path fading = north] (7.65,1)  to [out=60,in=-60] (7.825,2.75);
\draw [ultra thick, red, -stealth] (7.99,1.8)  to  (8,1.9);

\draw [ultra thick, red, -stealth] (7.9,4)  to  (7.89,4.1);
\draw [ultra thick, red, path fading = south] (8.1,3.25)  to [out=115,in=-115]  (8.15,5);
%\draw [ultra thick, red] (1.75,2.75) to (2.7,3.4);
\draw [ultra thick, red, path fading = north] (8.15,1)  to [out=65,in=-65] (8.3,2.75);
\draw [ultra thick, red, -stealth] (8.43,1.8)  to  (8.44,1.9);

\draw [ultra thick, red, -stealth] (8.44,4)  to  (8.43,4.1);
\draw [ultra thick, red, path fading = south] (8.6,3.25)  to [out=110,in=-110]  (8.65,5);
%\draw [ultra thick, red] (1.75,2.75) to (2.7,3.4);
\draw [ultra thick, red, path fading = north] (8.65,1)  to [out=70,in=-70] (8.775,2.75);
\draw [ultra thick, red, -stealth] (8.88,1.8)  to  (8.89,1.9);

\draw [ultra thick, red, -stealth] (8.99,4)  to  (8.98,4.1);
\draw [ultra thick, red, path fading = south] (9.1,3.25)  to [out=105,in=-105]  (9.15,5);
%\draw [ultra thick, red] (1.75,2.75) to (2.7,3.4);
\draw [ultra thick, red, path fading = north] (9.15,1)  to [out=75,in=-75] (9.25,2.75);
\draw [ultra thick, red, -stealth] (9.33,1.8)  to  (9.34,1.9);

%curved dashed left
\draw [ultra thick, dashed] (6.5,5.) -- (6.5,6);
\draw [ultra thick, dashed] (6.5,3)  to [out=135,in=-135] (6.5,5);
\draw [ultra thick, dashed] (6.5,1)  to [out=45,in=-45] (6.5,3);
\draw [ultra thick, dashed] (6.5,0.) -- (6.5,1);
\draw [ultra thick, dashed, path fading=east] (6.5,0) -- (7,0.3);
\draw [ultra thick, dashed, path fading=east] (6.5,6) -- (7,6.3);
\draw [ultra thick, dashed, path fading=east] (6.5,5) -- (7,5.3);
\draw [ultra thick, dashed, path fading=east] (6.1,4) -- (6.6,4.3);
\draw [ultra thick, dashed, path fading=east] (6.5,3) -- (7,3.3);
\draw [ultra thick, dashed, path fading=east] (7,2) -- (7.5,2.3);
\draw [ultra thick, dashed, path fading=east] (6.5,1) -- (7,1.3);

%curved dashed right vertical
\draw [ultra thick, dashed] (3.5,5) --(3.5,6);
\draw [ultra thick, dashed] (3.5,3) to [out=45,in=-90]  (4.25,3.875) to [out=90,in=-45]  (3.5,5);
\draw [ultra thick, dashed] (3.5,1) to [out=135,in=-90]  (2.75,2.125) to [out=90,in=-135]  (3.5,3);
\draw [ultra thick, dashed] (3.5,0) --(3.5,1);

%curved dashed right horizontal
\draw [ultra thick, dashed, path fading=east] (3.5,6) -- (4,6.3);
\draw [ultra thick, dashed, path fading=east] (3.5,5) -- (4,5.3);
\draw [ultra thick, dashed, path fading=east] (4.25,4) -- (4.75,4.3);
\draw [ultra thick, dashed, path fading=east] (3.5,3) -- (4,3.3);
\draw [ultra thick, dashed, path fading=east] (2.75,2) -- (3.25,2.3);
\draw [ultra thick, dashed, path fading=east] (3.5,1) -- (4,1.3);
\draw [ultra thick, dashed, path fading=east] (3.5,0) -- (4,0.3);

%bottom and top dashed right
\draw [ultra thick, dashed] (6.5,5) --(10,5);
\draw [ultra thick, dashed, path fading=east] (7.5,5) -- (8,5.3);
\draw [ultra thick, dashed, path fading=east] (8.5,5) -- (9,5.3);
\draw [ultra thick, dashed, path fading=east] (9.5,5) -- (10,5.3);
\draw [ultra thick, dashed] (6.5,1) --(10,1);
\draw [ultra thick, dashed, path fading=east] (7.5,1) -- (8,1.3);
\draw [ultra thick, dashed, path fading=east] (8.5,1) -- (9,1.3);
\draw [ultra thick, dashed, path fading=east] (9.5,1) -- (10,1.3);

\draw [<-] (0,6) -- (0,-0.3); 	%z
\draw [->] (0,1) -- (10,1);		%x
\draw [->] (0,1) -- (0.5,1.3);	%y

\small

\node [below] at (10,1) {$x$};
\node [left] at (0,6) {$z$};
\node [right] at (0.45,1.3) {$y$};

\large
\node [right] at (0.55,3) {$\rho_1$,$p_1$,$T_1$,$B_1$};
\node [right] at (3.85,3) {$\rho_0$,$p_0$,$T_0$,$B_0$};
\node [right] at (7.1,3) {$\rho_2$,$p_2$,$T_2$,$B_2$};
\node [left] at (-0.5,1) {$z=0$};
\node [left] at (-0.5,5) {$z=L$};
\node [below] at (3.5,-0.2) {$x=-x_0$};
\node [below] at (6.5,-0.2) {$x=x_0$};

%magnetic field end points
\fill[red] (2.25,5) circle (0.08cm);
\fill[red] (2.25,1) circle (0.08cm);
\fill[red] (0.35,5) circle (0.08cm);
\fill[red] (0.35,1) circle (0.08cm);
\fill[red] (4.25,5) circle (0.08cm);
\fill[red] (4.25,1) circle (0.08cm);
\fill[red] (5.2,5) circle (0.08cm);
\fill[red] (5.2,1) circle (0.08cm);
\fill[red] (5.85,5) circle (0.08cm);
\fill[red] (5.85,1) circle (0.08cm);
\fill[red] (7.15,5) circle (0.08cm);
\fill[red] (7.155,1) circle (0.08cm);
\fill[red] (7.65,5) circle (0.08cm);
\fill[red] (7.65,1) circle (0.08cm);
\fill[red] (8.15,5) circle (0.08cm);
\fill[red] (8.15,1) circle (0.08cm);
\fill[red] (8.65,5) circle (0.08cm);
\fill[red] (8.65,1) circle (0.08cm);
\fill[red] (9.15,5) circle (0.08cm);
\fill[red] (9.15,1) circle (0.08cm);
\end{tikzpicture}
}
\caption{Same as Figure \ref{fig:standing_saus_fund} but for the first harmonic.}
\label{fig:standing_saus_first}
  \end{minipage}
\end{figure}

\begin{figure}
	\begin{minipage}[t]{8.5cm}
		\includegraphics[scale=0.32, trim={1.5cm 1cm 2cm 2.5cm},clip]{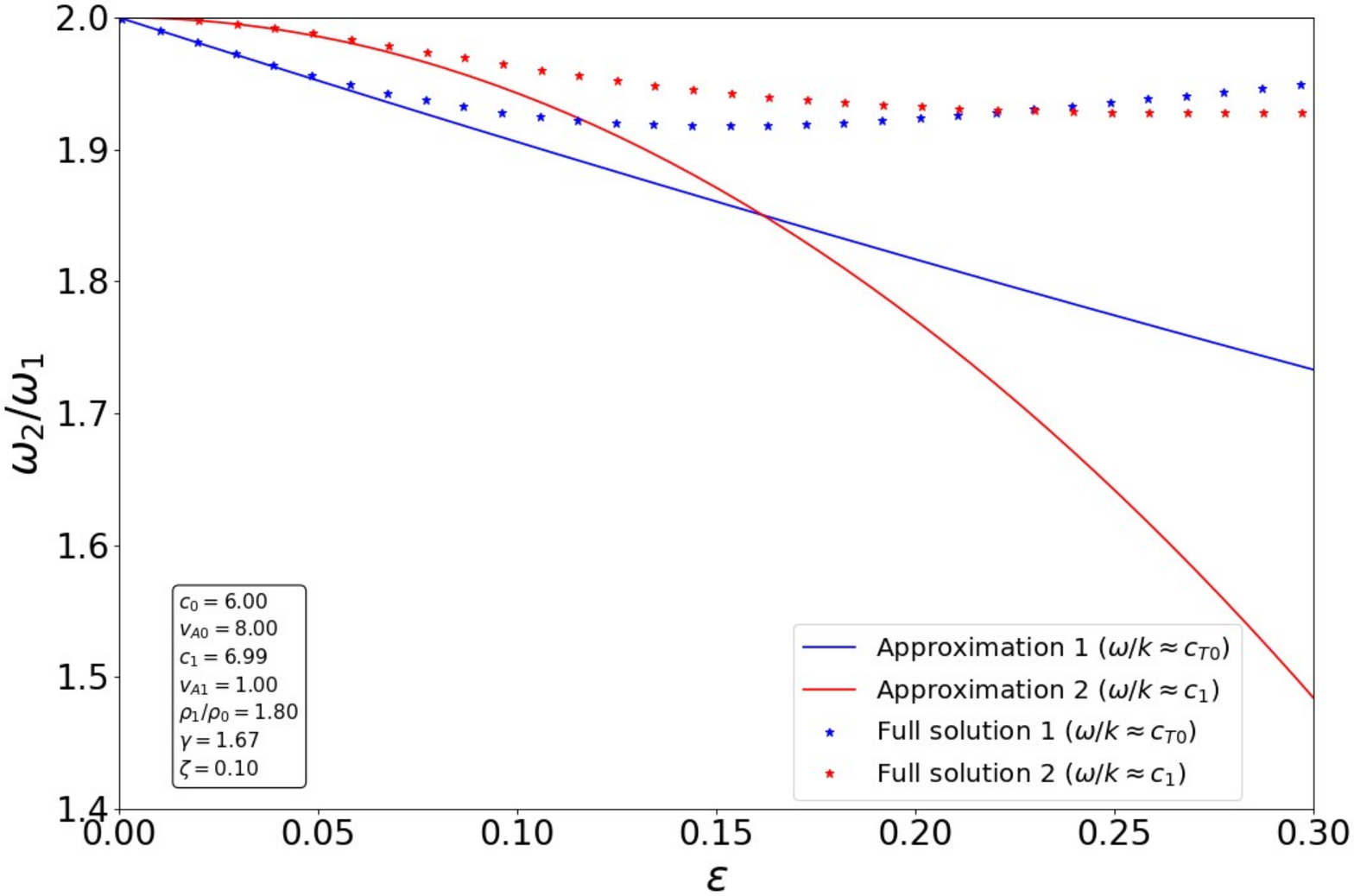}
		\caption{The ratio of the frequencies of the first harmonic to the fundamental mode of the quasi-sausage waves (Solution 1 given is by Equation (\ref{fi3}) and Solution 2 by Equation \ref{fi7}), as a function of $\varepsilon$ ranging from $0$ to $0.3$, with $\zeta=0.1$ fixed. The other relevant quantities are given in the inlet of the figure. Note that due to the isothermal assumption $c_1=c_2$.}
		\label{fig:ratio_epsilon}
	\end{minipage}
\hspace{0.5cm}
		\begin{minipage}[t]{8.5cm}
		\includegraphics[scale=0.32, trim={1cm 1cm 2cm 2.5cm},clip]{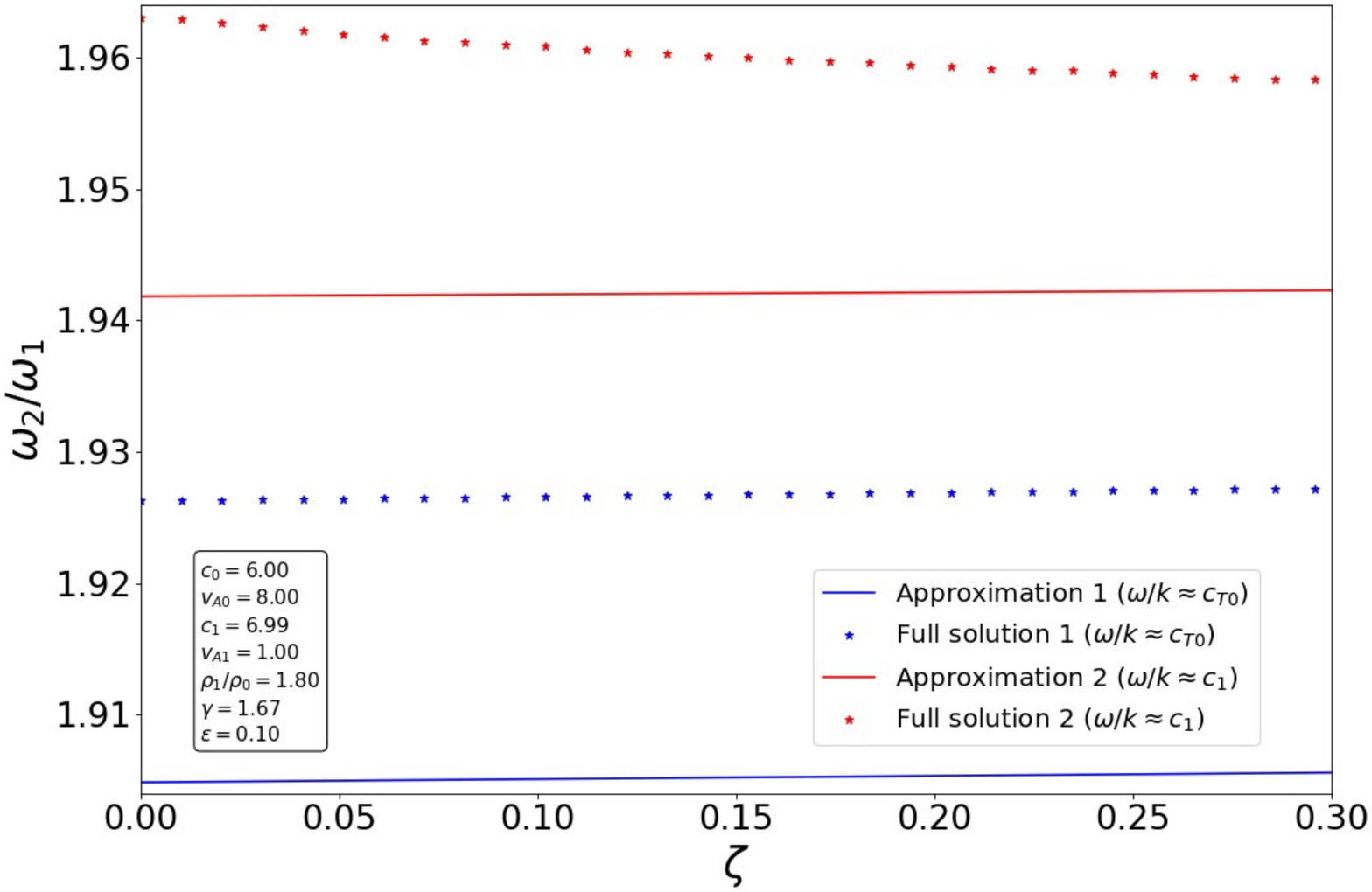}
		\caption{Same as in Figure \ref{fig:ratio_epsilon}, but the ratio is plotted as a function of $\zeta$ ranging from $0$ to $0.3$, with $\varepsilon=0.1$ fixed.}
		\label{fig:ratio_zeta}
	\end{minipage}
\end{figure}

\subsection{Quasi-kink Modes} \label{QK1}
Let us define the kink speed, $v_k$ by

\begin{equation}
v_k^2 = \frac{\rho_1v_{A1}^2+\rho_2v_{A2}^2}{\rho_1+\rho_2}\approx v_{A1}^2\left(1+\frac{\zeta}{2}\right),
\label{fi8}
\end{equation}
where the approximation is written using Equation (\ref{dr8}) and keeping only first-order terms in $\zeta$. Then, by examining Equation (\ref{dr11}), we see there is a solution given by $\omega^2 = k^2v_k^2(1+\alpha_1)$, where $\alpha_1 \sim O(\varepsilon^2)$. The quantity $v_k^2$ contains an asymmetry term as shown in Equation (\ref{fi8}), and it would be possible to make a similar expansion about both $k^2v_{A1}^2$ or $k^2v_{A2}^2$, as they both differ from $k^2v_k^2$ by a factor of $\zeta/2$ to first-order. However, using the kink speed is much more natural, as when the slab width is reduced to zero, we must recover the solution given in \cite{rob-81-a} for waves at a single interface, and this solution is $\omega^2=k^2v_k^2$. 
The assumption that $m_1^2$, $m_2^2>0$, along with our desire to make analytical progress, cause a change in the orderings of $\zeta$ and $\varepsilon$. More specifically, take $\varepsilon^3 \ll \zeta \ll \varepsilon^2 \ll \varepsilon \ll 1$. A solution can be derived of the form

\begin{equation}
\omega^2 \approx k^2v_k^2(1+A\varepsilon^2 +B\frac{\zeta^2}{\varepsilon^2}), \quad \text{where} \quad A,B \sim O(1).
\label{fi9}
\end{equation}
Unfortunately, the limitations of the method used here are evident, and the term of size $\zeta^2/\varepsilon^2$ is of comparable size to the terms ignored in order to decouple the full dispersion relation. Consequently, a valid solution that includes the first explicit term due to asymmetry cannot be determined. From here, it would be possible to take a further ordering, say $\varepsilon^4 \ll \zeta \ll \varepsilon^3 \ll 1$, however, this work aims to analyse the effect of asymmetry, and the smaller we take the asymmetry to be, the less significant its effect. A solution can be found up to the $\varepsilon^2$-term as follows:

\begin{equation}
\omega_n^2 \approx \frac{n^2\pi^2v_k^2}{L^2}\left(1-\varepsilon^2\frac{n^2\pi^2\rho_0^2(v_k^2-v_{A0}^2)^2(v_k^2-c_1^2)}{v_k^6\rho_1^2}\right).
\label{fi10}
\end{equation}
Expression (\ref{fi10}) is a similar result to Equation (18a) in \cite{rob-82}, however, there is a small error in the paper by Edwin and Roberts. More precisely, in Equation (18a) the factor $(1-v_A^2/v_{Ae}^2)$ should have been squared. Care must be taken when using this solution as the method used in the derivation involves squaring terms, and so the solution only satisfies the decoupled dispersion relation Equation (\ref{dr11}) when the correct square-root is taken for a given ordering of characteristic speeds.

Let us now define the average external tube speed, $c_{TA}$ by:

\begin{equation}
c_{TA}^2 = \frac{c_{T1}^2+c_{T2}^2}{2} \approx c_{T1}^2\left(1+\frac{\zeta c_{T1}^2}{2v_{A1}^2} \right),
\label{fi11}
\end{equation}
where the approximation is written using Equations (\ref{dr5}) and (\ref{dr8}) and keeping only first-order terms in $\zeta$. Then, there is another solution to Equation (\ref{dr11}), namely given by $\omega^2 = k^2c_{TA}^2(1+\alpha_2)$, where $\alpha_2 \sim O(\varepsilon^2)$. To make analytical progress we, again, assume $\varepsilon^3 \ll \zeta \ll \varepsilon^2 \ll \varepsilon \ll 1$, giving

\begin{equation}
\omega_n^2 \approx \frac{n^2\pi^2c_{TA}^2}{L^2}\left(1-\varepsilon^2\frac{n^2\pi^2\rho_0^2(c_{TA}^2-v_{A0}^2)^2(c_1^2-c_{TA}^2)}{v_{A1}^4c_{TA}^2\rho_1^2}\right),
\label{fi12}
\end{equation}
valid only when $c_{TA}>v_{A0}$. Additionally, $c_{TA}<c_0$ must be taken for reasons discussed in the introduction to Section \ref{Frequencies of the Standing Harmonic Modes}. As with the first quasi-kink solution, an expression involving the first explicit asymmetry term cannot be determined (apart from the asymmetry contained in $c_{TA}^2$). This is similar to the expression obtained in Equation (18b) in \cite{rob-82}, and we can reduce Equation (\ref{fi12}) to that equation by taking $\zeta=0$.

Both solutions for quasi-kink waves do not display any explicit dependence on the magnetic asymmetry parameter $\zeta$. However, defining the speeds $v_k$ and $c_{TA}$, the asymmetry dependence is contained in these quantities and allow for an expansion of the solution such that the term of size $\zeta$ disappears. Therefore the effects of magnetic asymmetry are present in the speed that the expansion is about.

The frequency ratio of the first harmonic eigenmode to the fundamental mode is not calculated for either of the quasi-kink frequency solutions. This is because the study is concerned with analysing magnetic asymmetry, and as the eigenfrequency solutions contain no explicit asymmetry terms, the frequency ratio will not contain terms explicitly involving the magnetic asymmetry parameter $\zeta$ and is therefore not of interest.

None of the eigenfrequencies for the quasi-kink modes can be compared to the solutions given in \cite{ox-20}. This is because the solutions found here are derived under the assumption that the external magnetic field is not insignificant when compared to the internal magnetic field, and if we were to try and reduce the size of the external field $B_1$ to zero, we would have to make a separate analysis. Doing so would recover a similar solution to Equation (31) in \cite{ox-20}, however, that analysis under limiting conditions is not included. The main reason for this is that our study is focused on the effects of the external magnetism on the standing waves.

Illustrations of the fundamental and first harmonic standing quasi-kink waves are given in Figures \ref{fig:standing_kink_fund} and \ref{fig:standing_kink_first}.

\begin{figure}[h!]
\hspace{0.2cm}
  \begin{minipage}[t]{0.45\textwidth}
\centering
\scalebox{0.7}[0.7]{
\begin{tikzpicture}
%middle
\path [fill=lightgray, opacity=0.6] (3.5,5) -- (3.5,6) -- (6.5,6) -- (6.5,5) -- (3.5,5);
\path [fill=lightgray, opacity=0.6] (3.5,0) -- (3.5,1) -- (6.5,1) -- (6.5,0) -- (3.5,0);
%\path [fill=lightgray, opacity=0.6] (3.5,1) to [out=135, in=-135] (3.5,5) to (6.5,5) to [out=-45, in=90] (7.25, 3.875) to [out=-90, in=45] (6.5,3) to [out=-135, in=90] (5.75,2.125) to [out=-90, in=135] (6.5,1) to (3.5,1);
\path [fill=lightgray, opacity=0.6] (3.5,1) to [out=115, in=-115] (3.5,5) to  (6.5,5) to [out=-140, in=140] (6.5,1) to (3.5,1);

%curved surfaces middle
\shade[left color=darkgray,right color=white, opacity=0.99] (6.5,0) -- (6.5,1) -- (7,1.3) -- (7,0.3) -- (6.5,0);
\shade[left color=darkgray,right color=white, opacity=0.99] (6.5,5) -- (6.5,6) -- (7,6.3) -- (7,5.3) -- (6.5,5);
%\shade[left color=darkgray,right color=white, opacity=0.99] (6.5,1) to [out=135, in=-90] (5.75,2.125) to [out=90, in=-135] (6.5,3) to [out=45, in=-90] (7.25, 3.875) to [out=90, in=-45] (6.5,5) to (7,5.3) to [out=-45, in=90] (7.75,4.175) to [out=-90, in=45] (7,3.3) to [out=-135, in=90] (6.25,2.425) to [out=-90, in=135] (7,1.3) to (6.5,1);
\shade[left color=darkgray,right color=white, opacity=0.99] (6.5,1) to [out=140, in=-140] (6.5,5) to (7,5.3) to [out=-140, in=140] (7,1.3) to (6.5,1);
\shade[top color=white,bottom color=lightgray, opacity=0.7] (-0.5,1) -- (0.,1.3) -- (4,1.3) -- (3.5,1) -- (-0.5,1);
\shade[bottom color=lightgray,top color=white, opacity=0.45] (3.5,1) -- (6.5,1) -- (6.5,1.3) -- (4.,1.3) -- (3.5,1);
\shade[left color=darkgray,right color=white, opacity=0.3] (3.5,1) to [out=115, in=-115] (3.5,5) to (4,5.3) to [out=-115, in=115] (4,1.3) to (3.5,1);

%middle top and bottom
\shade[top color=lightgray,bottom color=white, opacity=0.45] (3.5,0) -- (6.5,0) -- (6.5,-0.3) -- (3.5,-0.3) -- (3.5,0);
\shade[top color=white,bottom color=lightgray, opacity=0.45] (3.5,6) -- (4,6.3) -- (7,6.3) -- (6.5,6) -- (3.5,6);
\shade[top color=white,bottom color=lightgray, opacity=0.45] (3.5,5) -- (4,5.3) -- (7,5.3) -- (6.5,5) -- (3.5,5);
\shade[left color=lightgray,right color=white, opacity=0.99] (6.5,0) -- (6.5,-0.3) -- (7,0) -- (7,0.3) -- (6.5,0);

%curved surface right + very top + very bottom right
\shade[top color=lightgray,bottom color=white, opacity=0.8] (6.5,0) -- (10,0) -- (10,-0.3) -- (6.5,-0.3) -- (6.5,0);
\shade[top color=white,bottom color=lightgray, opacity=0.8] (6.5,6) -- (7,6.3) -- (10.5,6.3) -- (10,6) -- (6.5,6);
\shade[right color=white,left color=lightgray, opacity=0.8] (10,-0.) -- (10,1) -- (10.5,1.3) -- (10.5,0.3) -- (10,-0.);
\shade[bottom color=white,top color=lightgray, opacity=0.3] (10,-0.) -- (10,1) -- (10.5,1.3) -- (10.5,0.3) -- (10,-0.);
\shade[right color=white, left color=lightgray, opacity=0.8] (10,5.) -- (10,6) -- (10.5,6.3) -- (10.5,5.3) -- (10,5.);
\shade[top color=white, bottom color=lightgray, opacity=0.8] (10,5.) -- (10,6) -- (10.5,6.3) -- (10.5,5.3) -- (10,5.);
%\shade[left color=lightgray,right color=white, opacity=0.8] (10,1) to [out=135,in=-90] (9.25,2.125) to [out=90,in=-135] (10,3.) to [out=45, in=-90] (10.75,3.875) to [out=90, in=-45] (10,5) to (10.5, 5.3) to [out=-45, in=90] (11.25, 4.175) to [out=-90, in=45] (10.5,3.3) to [out=-135, in=90] (9.75, 2.425) to [out=-90, in=135] (10.5,1.3);
\shade[top color=lightgray,bottom color=white, opacity=0.3] (10,-0.3) -- (10,0.) -- (10.5,0.3) -- (10.5,-0.) -- (10,-0.3);
\shade[left color=lightgray, right color=white, opacity=0.3] (10,-0.3) -- (10,0.) -- (10.5,0.3) -- (10.5,-0.) -- (10,-0.3);
\shade[left color=lightgray,right color=white, opacity=0.9] (10,1) -- (9.5,2) -- (10,3) -- (9.5,4) -- (10,5) -- (10.5,5.3) -- (10,4.3) -- (10.5,3.3) -- (10,2.3) -- (10.5,1.3) -- (10,1);

%curved surface left
\shade[top color=lightgray,bottom color=white, opacity=0.3] (-0.5,0) -- (3.5,0) -- (3.5,-0.3) -- (-0.5,-0.3) -- (-0.5,0);
\shade[top color=white,bottom color=lightgray, opacity=0.3] (-0.5,6) -- (0.,6.3) -- (4,6.3) -- (3.5,6) -- (-0.5,6);
\shade[top color=white,bottom color=lightgray, opacity=0.7] (-0.5,5) -- (0.,5.3) -- (4,5.3) -- (3.5,5) -- (-0.5,5);
%\shade[right color=white,left color=lightgray, opacity=0.5] (-0.5,1) to [out=135, in=-135] (-0.5,3) to [out=45, in=-45] (-0.5,5) to (0,5.3) to [out=-45, in=45] (0,3.3) to [out=-135, in=135] (0,1.3) to (-0.5,1);
\shade[right color=white,left color=lightgray, opacity=0.5] (0,1.3) to (-0.5,2.3) to  (0,3.3) to (-0.5,4.3) to (0,5.3) to (0,5.3) to (0,3.3) to  (0,1.3) to (-0.5,1.3);

%top and bottom corners right
\shade[top color=white,bottom color=black, opacity=0.9] (6.5,1) -- (7.,1.3) -- (10.5,1.3) -- (10,1) -- (6.5,1);
\shade[top color=white,bottom color=black, opacity=0.9] (6.5,5) -- (7.,5.3) -- (10.5,5.3) -- (10,5) -- (6.5,5);

%top dashed left
\draw [ultra thick, dashed] (-0.5,5.) -- (3.5,5);
\draw [ultra thick, dashed, path fading = east] (-0.5,5.) -- (0,5.3);
\draw [ultra thick, dashed, path fading = east] (0.5,5.) -- (1,5.3);
\draw [ultra thick, dashed, path fading = east] (1.5,5.) -- (2,5.3);
\draw [ultra thick, dashed, path fading = east] (2.5,5.) -- (3,5.3);

%bottom dashed left
\draw [ultra thick, dashed] (-0.5,1.) -- (3.5,1);
\draw [ultra thick, dashed, path fading = east] (-0.5,1.) -- (0,1.3);
\draw [ultra thick, dashed, path fading = east] (0.5,1.) -- (1,1.3);
\draw [ultra thick, dashed, path fading = east] (1.5,1.) -- (2,1.3);
\draw [ultra thick, dashed, path fading = east] (2.5,1.) -- (3,1.3);

%curved dashed left
\draw [ultra thick, dashed] (3.5,5.) -- (3.5,6);
\draw [ultra thick, dashed] (3.5,1)  to [out=115,in=-115] (3.5,5);
%\draw [ultra thick, dashed] (3.5,1)  to [out=135,in=-135] (3.5,3);
\draw [ultra thick, dashed] (3.5,0.) -- (3.5,1);
\draw [ultra thick, dashed, path fading=east] (3.5,0) -- (4,0.3);
\draw [ultra thick, dashed, path fading=east] (3.5,6) -- (4,6.3);
\draw [ultra thick, dashed, path fading=east] (3.5,5) -- (4,5.3);
\draw [ultra thick, dashed, path fading=east] (3.18,4) -- (3.68,4.3);
\draw [ultra thick, dashed, path fading=east] (3.05,3) -- (3.55,3.3);
\draw [ultra thick, dashed, path fading=east] (3.1,2) -- (3.6,2.3);
\draw [ultra thick, dashed, path fading=east] (3.5,1) -- (4,1.3);

%top and bottom dashed middle
\draw [ultra thick, dashed] (3.5,5) --(6.5,5);
\draw [ultra thick, dashed, path fading=east] (4.5,5) -- (5,5.3);
\draw [ultra thick, dashed, path fading=east] (5.5,5) -- (6,5.3);
\draw [ultra thick, dashed] (3.5,1) --(6.5,1);
\draw [ultra thick, dashed, path fading=east] (4.5,1) -- (5,1.3);
\draw [ultra thick, dashed, path fading=east] (5.5,1) -- (6,1.3);

%1 filling
\shade[top color=lightgray, bottom color=lightgray, opacity=0.3] (-0.5,1.) to (-0.5,0.) to (3.5,0) to (3.5,1) to (-0.5,1);
\shade[top color=lightgray, bottom color=lightgray, opacity=0.3] (-0.5,1) to (-1,2) to (-0.5,3) to (-1,4) to (-0.5,5) to (3.5,5) to [out=-115, in=115] (3.5,1) to (-0.5,1);
%\shade[top color=lightgray, bottom color=lightgray, opacity=0.3] (-0.5,5.) to (3.5,5.) to [out=-45,in=45] (3.5,3) to (-0.5,3) to (-1,4) to (-0.5,5);
\shade[top color=lightgray, bottom color=lightgray, opacity=0.3] (-0.5,5.) to (-0.5,6.) to (3.5,6) to (3.5,5) to (-0.5,5);

%2 filling
\shade[top color=lightgray, bottom color=lightgray, opacity=0.9] (6.5,1.) to (6.5,0.) to (10.,0) to (10.,1) to (6.5,1);
%\shade[top color=lightgray, bottom color=lightgray, opacity=0.9] (6.5,1) to [out=135,in=-90] (5.75,2.125)
to [out=90,in=-135] (6.5,3.) to (10.,3) to (9.5,2) to (10.,1) to (6.5,1);
\shade[top color=lightgray, bottom color=lightgray, opacity=0.9] (6.5,5.) to (10.,5.) to (9.5,4) to (10.,3) to (9.5,2) to (10.,1) to (6.5,1) to [out=140,in=-140]  (6.5,5);
\shade[top color=lightgray, bottom color=lightgray, opacity=0.9] (6.5,5.) to (6.5,6.) to (10.,6) to (10.,5) to (6.5,5);

%\path [fill=lightgray, opacity=0.9] (6.5,1) -- (6.5,5) -- (10,5) -- (9.5,4) -- (10,3) -- (9.5,2) -- (10,1) -- (6.5,1);

% magnetic field arrows middle
\draw [ultra thick, red, -stealth] (3.7,4)  to  (3.73,4.1);
\draw [ultra thick, red, path fading = south] (3.6,3.25)  to [out=85,in=-135]  (4.25,5);
\draw [ultra thick, red, path fading = north] (4.25,1)  to [out=135,in=-85] (3.6,2.75);
\draw [ultra thick, red, -stealth] (3.76,1.8)  to  (3.73,1.9);

\draw [ultra thick, red, -stealth] (4.39,4)  to  (4.42,4.1);
\draw [ultra thick, red, path fading = south] (4.25,3.25)  to [out=85,in=-135]  (5.,5);
\draw [ultra thick, red, path fading = north] (5.,1)  to [out=135,in=-85] (4.25,2.75);
\draw [ultra thick, red, -stealth] (4.45,1.8)  to  (4.42,1.9);

\draw [ultra thick, red, -stealth] (5.02,4)  to  (5.06,4.1);
\draw [ultra thick, red, path fading = south] (4.85,3.25)  to [out=85,in=-135]  (5.75,5);
\draw [ultra thick, red, path fading = north] (5.75,1)  to [out=135,in=-85] (4.85,2.75);
\draw [ultra thick, red, -stealth] (5.11,1.8)  to  (5.06,1.9);

% magnetic field arrows left
\draw [ultra thick, red, -stealth] (2.2,4)  to  (2.23,4.1);
\draw [ultra thick, red, path fading = south] (2.1,3.25)  to [out=85,in=-135]  (2.75,5);
\draw [ultra thick, red, path fading = north] (2.75,1)  to [out=135,in=-85] (2.1,2.75);
\draw [ultra thick, red, -stealth] (2.26,1.8)  to  (2.23,1.9);

\draw [ultra thick, red, -stealth] (0.7,4)  to  (0.73,4.1);
\draw [ultra thick, red, path fading = south] (0.65,3.25)  to [out=90,in=-115]  (1.05,5);
%\draw [ultra thick, red] (0.65,2.75) to (0.65,3.25);
\draw [ultra thick, red, path fading = north] (1.05,1)  to [out=115,in=-90] (0.65,2.75);
\draw [ultra thick, red, -stealth] (0.76,1.8)  to  (0.73,1.9);

% magnetic field arrows right

\draw [ultra thick, red, -stealth] (6.62,4)  to  (6.66,4.1);
\draw [ultra thick, red] (6.45,3)  to [out=90,in=-135]  (7.35,5);
\draw [ultra thick, red] (7.35,1)  to [out=135,in=-90] (6.45,3);
\draw [ultra thick, red, -stealth] (6.71,1.8)  to  (6.66,1.9);

\draw [ultra thick, red, -stealth] (7.2,4)  to  (7.24,4.1);
\draw [ultra thick, red, path fading = south] (7.05,3.25)  to [out=85,in=-130]  (7.85,5);
\draw [ultra thick, red, path fading = north] (7.85,1)  to [out=130,in=-85] (7.05,2.75);
\draw [ultra thick, red, -stealth] (7.29,1.8)  to  (7.24,1.9);

\draw [ultra thick, red, -stealth] (7.78,4)  to  (7.82,4.1);
\draw [ultra thick, red, path fading = south] (7.65,3.25)  to [out=85,in=-125]  (8.35,5);
\draw [ultra thick, red, path fading = north] (8.35,1)  to [out=125,in=-85] (7.65,2.75);
\draw [ultra thick, red, -stealth] (7.87,1.8)  to  (7.82,1.9);

\draw [ultra thick, red, -stealth] (8.35,4)  to  (8.39,4.1);
\draw [ultra thick, red, path fading = south] (8.2,3.25)  to [out=85,in=-120]  (8.85,5);
\draw [ultra thick, red, path fading = north] (8.85,1)  to [out=120,in=-85] (8.25,2.75);
\draw [ultra thick, red, -stealth] (8.44,1.8)  to  (8.4,1.9);

\draw [ultra thick, red, -stealth] (8.95,4)  to  (8.97,4.1);
\draw [ultra thick, red, path fading = south] (8.85,3.25)  to [out=85,in=-115]  (9.35,5);
%\draw [ultra thick, red] (8.85,2.55) to [out=95,in=-95] (8.85,3.45);
\draw [ultra thick, red, path fading = north] (9.35,1)  to [out=115,in=-85] (8.85,2.75);
\draw [ultra thick, red, -stealth] (9.02,1.8)  to  (8.99,1.9);

%curved dashed right vertical
\draw [ultra thick, dashed] (6.5,5) --(6.5,6);
\draw [ultra thick, dashed] (6.5,1) to [out=140,in=-140]  (6.5,5);
%\draw [ultra thick, dashed] (6.5,1) to [out=135,in=-90]  (5.75,2.125) to [out=90,in=-135]  (6.5,3);
\draw [ultra thick, dashed] (6.5,0) --(6.5,1);

%curved dashed right horizontal
\draw [ultra thick, dashed, path fading=east] (6.5,6) -- (7,6.3);
\draw [ultra thick, dashed, path fading=east] (6.5,5) -- (7,5.3);
\draw [ultra thick, dashed, path fading=east] (5.8,4) -- (6.3,4.3);
\draw [ultra thick, dashed, path fading=east] (5.65,3) -- (6.15,3.3);
\draw [ultra thick, dashed, path fading=east] (5.75,2) -- (6.25,2.3);
\draw [ultra thick, dashed, path fading=east] (6.5,1) -- (7,1.3);
\draw [ultra thick, dashed, path fading=east] (6.5,0) -- (7,0.3);

%bottom and top dashed right
\draw [ultra thick, dashed] (6.5,5) --(10,5);
\draw [ultra thick, dashed, path fading=east] (7.5,5) -- (8,5.3);
\draw [ultra thick, dashed, path fading=east] (8.5,5) -- (9,5.3);
\draw [ultra thick, dashed, path fading=east] (9.5,5) -- (10,5.3);
\draw [ultra thick, dashed] (6.5,1) --(10,1);
\draw [ultra thick, dashed, path fading=east] (7.5,1) -- (8,1.3);
\draw [ultra thick, dashed, path fading=east] (8.5,1) -- (9,1.3);
\draw [ultra thick, dashed, path fading=east] (9.5,1) -- (10,1.3);

\draw [<-] (0,6) -- (0,-0.3); 	%z
\draw [->] (0,1) -- (10,1);		%x
\draw [->] (0,1) -- (0.5,1.3);	%y

\small
\node [below] at (10,1) {$x$};
\node [left] at (0,6) {$z$};
\node [right] at (0.45,1.3) {$y$};

\large
\node [right] at (0.45,3) {$\rho_1$,$p_1$,$T_1$,$B_1$};
\node [right] at (3.35,3) {$\rho_0$,$p_0$,$T_0$,$B_0$};
\node [right] at (6.8,3) {$\rho_2$,$p_2$,$T_2$,$B_2$};
\node [below] at (3.5,-0.2) {$x=-x_0$};
\node [below] at (6.5,-0.2) {$x=x_0$};
\node [left] at (-0.5,1) {$z=0$};
\node [left] at (-0.5,5) {$z=L$};
%magnetic field end points

\fill[red] (1.05,5) circle (0.08cm);
\fill[red] (1.05,1) circle (0.08cm);
\fill[red] (2.75,5) circle (0.08cm);
\fill[red] (2.75,1) circle (0.08cm);
\fill[red] (4.25,5) circle (0.08cm);
\fill[red] (4.25,1) circle (0.08cm);
\fill[red] (5,5) circle (0.08cm);
\fill[red] (5,1) circle (0.08cm);
\fill[red] (5.75,5) circle (0.08cm);
\fill[red] (5.75,1) circle (0.08cm);
\fill[red] (7.35,5) circle (0.08cm);
\fill[red] (7.35,1) circle (0.08cm);
\fill[red] (7.85,5) circle (0.08cm);
\fill[red] (7.85,1) circle (0.08cm);
\fill[red] (8.35,5) circle (0.08cm);
\fill[red] (8.35,1) circle (0.08cm);
\fill[red] (8.85,5) circle (0.08cm);
\fill[red] (8.85,1) circle (0.08cm);
\fill[red] (9.35,5) circle (0.08cm);
\fill[red] (9.35,1) circle (0.08cm);
\end{tikzpicture}
}
\caption{Illustration of a fundamental standing quasi-kink mode oscillation in the slab embedded in a magnetically asymmetric environment.}
\label{fig:standing_kink_fund}
  \end{minipage}
 \hspace{0.5cm}
  \begin{minipage}[t]{0.45\textwidth}
\centering
\scalebox{0.7}[0.7]{
\begin{tikzpicture}
%middle
\path [fill=lightgray, opacity=0.6] (3.5,5) -- (3.5,6) -- (6.5,6) -- (6.5,5) -- (3.5,5);
\path [fill=lightgray, opacity=0.6] (3.5,0) -- (3.5,1) -- (6.5,1) -- (6.5,0) -- (3.5,0);
\path [fill=lightgray, opacity=0.6] (3.5,1) to [out=135, in=-135] (3.5, 3.)  to [out=45, in=-45] (3.5,5) to (6.5,5) to [out=-45, in=90] (7.25, 3.875) to [out=-90, in=45] (6.5,3) to [out=-135, in=90] (5.75,2.125) to [out=-90, in=135] (6.5,1) to (3.5,1);

%curved surfaces middle
\shade[left color=darkgray,right color=white, opacity=0.99] (6.5,0) -- (6.5,1) -- (7,1.3) -- (7,0.3) -- (6.5,0);
\shade[left color=darkgray,right color=white, opacity=0.99] (6.5,5) -- (6.5,6) -- (7,6.3) -- (7,5.3) -- (6.5,5);
\shade[left color=darkgray,right color=white, opacity=0.99] (6.5,1) to [out=135, in=-90] (5.75,2.125) to [out=90, in=-135] (6.5,3) to [out=45, in=-90] (7.25, 3.875) to [out=90, in=-45] (6.5,5) to (7,5.3) to [out=-45, in=90] (7.75,4.175) to [out=-90, in=45] (7,3.3) to [out=-135, in=90] (6.25,2.425) to [out=-90, in=135] (7,1.3) to (6.5,1);
\shade[top color=white,bottom color=lightgray, opacity=0.7] (-0.5,1) -- (0.,1.3) -- (4,1.3) -- (3.5,1) -- (-0.5,1);
\shade[bottom color=lightgray,top color=white, opacity=0.45] (3.5,1) -- (6.5,1) -- (6.5,1.3) -- (4.,1.3) -- (3.5,1);
\shade[left color=darkgray,right color=white, opacity=0.3] (3.5,1) to [out=135, in=-135] (3.5,3) to [out=45, in=-45] (3.5,5) to (4,5.3) to [out=-45, in=45] (4,3.3) to [out=-135, in=135] (4,1.3) to (3.5,1);

%middle top and bottom
\shade[top color=lightgray,bottom color=white, opacity=0.45] (3.5,0) -- (6.5,0) -- (6.5,-0.3) -- (3.5,-0.3) -- (3.5,0);
\shade[top color=white,bottom color=lightgray, opacity=0.45] (3.5,6) -- (4,6.3) -- (7,6.3) -- (6.5,6) -- (3.5,6);
\shade[top color=white,bottom color=lightgray, opacity=0.45] (3.5,5) -- (4,5.3) -- (7,5.3) -- (6.5,5) -- (3.5,5);
\shade[left color=lightgray,right color=white, opacity=0.99] (6.5,0) -- (6.5,-0.3) -- (7,0) -- (7,0.3) -- (6.5,0);

%curved surface right + very top + very bottom right
\shade[top color=lightgray,bottom color=white, opacity=0.8] (6.5,0) -- (10,0) -- (10,-0.3) -- (6.5,-0.3) -- (6.5,0);
\shade[top color=white,bottom color=lightgray, opacity=0.8] (6.5,6) -- (7,6.3) -- (10.5,6.3) -- (10,6) -- (6.5,6);
\shade[right color=white,left color=lightgray, opacity=0.8] (10,-0.) -- (10,1) -- (10.5,1.3) -- (10.5,0.3) -- (10,-0.);
\shade[bottom color=white,top color=lightgray, opacity=0.3] (10,-0.) -- (10,1) -- (10.5,1.3) -- (10.5,0.3) -- (10,-0.);
\shade[right color=white, left color=lightgray, opacity=0.8] (10,5.) -- (10,6) -- (10.5,6.3) -- (10.5,5.3) -- (10,5.);
\shade[top color=white, bottom color=lightgray, opacity=0.8] (10,5.) -- (10,6) -- (10.5,6.3) -- (10.5,5.3) -- (10,5.);
%\shade[left color=lightgray,right color=white, opacity=0.8] (10,1) to [out=135,in=-90] (9.25,2.125) to [out=90,in=-135] (10,3.) to [out=45, in=-90] (10.75,3.875) to [out=90, in=-45] (10,5) to (10.5, 5.3) to [out=-45, in=90] (11.25, 4.175) to [out=-90, in=45] (10.5,3.3) to [out=-135, in=90] (9.75, 2.425) to [out=-90, in=135] (10.5,1.3);
\shade[top color=lightgray,bottom color=white, opacity=0.3] (10,-0.3) -- (10,0.) -- (10.5,0.3) -- (10.5,-0.) -- (10,-0.3);
\shade[left color=lightgray, right color=white, opacity=0.3] (10,-0.3) -- (10,0.) -- (10.5,0.3) -- (10.5,-0.) -- (10,-0.3);
\shade[left color=lightgray,right color=white, opacity=0.9] (10,1) -- (9.5,2) -- (10,3) -- (9.5,4) -- (10,5) -- (10.5,5.3) -- (10,4.3) -- (10.5,3.3) -- (10,2.3) -- (10.5,1.3) -- (10,1);

%curved surface left
\shade[top color=lightgray,bottom color=white, opacity=0.3] (-0.5,0) -- (3.5,0) -- (3.5,-0.3) -- (-0.5,-0.3) -- (-0.5,0);
\shade[top color=white,bottom color=lightgray, opacity=0.3] (-0.5,6) -- (0.,6.3) -- (4,6.3) -- (3.5,6) -- (-0.5,6);
\shade[top color=white,bottom color=lightgray, opacity=0.7] (-0.5,5) -- (0.,5.3) -- (4,5.3) -- (3.5,5) -- (-0.5,5);
%\shade[right color=white,left color=lightgray, opacity=0.5] (-0.5,1) to [out=135, in=-135] (-0.5,3) to [out=45, in=-45] (-0.5,5) to (0,5.3) to [out=-45, in=45] (0,3.3) to [out=-135, in=135] (0,1.3) to (-0.5,1);
\shade[right color=white,left color=lightgray, opacity=0.5] (0,1.3) to (-0.5,2.3) to  (0,3.3) to (-0.5,4.3) to (0,5.3) to (0,5.3) to (0,3.3) to  (0,1.3) to (-0.5,1.3);

%top and bottom corners right
\shade[top color=white,bottom color=black, opacity=0.9] (6.5,1) -- (7.,1.3) -- (10.5,1.3) -- (10,1) -- (6.5,1);
\shade[top color=white,bottom color=black, opacity=0.9] (6.5,5) -- (7.,5.3) -- (10.5,5.3) -- (10,5) -- (6.5,5);

%top dashed left
\draw [ultra thick, dashed] (-0.5,5.) -- (3.5,5);
\draw [ultra thick, dashed, path fading = east] (-0.5,5.) -- (0,5.3);
\draw [ultra thick, dashed, path fading = east] (0.5,5.) -- (1,5.3);
\draw [ultra thick, dashed, path fading = east] (1.5,5.) -- (2,5.3);
\draw [ultra thick, dashed, path fading = east] (2.5,5.) -- (3,5.3);

%bottom dashed left
\draw [ultra thick, dashed] (-0.5,1.) -- (3.5,1);
\draw [ultra thick, dashed, path fading = east] (-0.5,1.) -- (0,1.3);
\draw [ultra thick, dashed, path fading = east] (0.5,1.) -- (1,1.3);
\draw [ultra thick, dashed, path fading = east] (1.5,1.) -- (2,1.3);
\draw [ultra thick, dashed, path fading = east] (2.5,1.) -- (3,1.3);

%curved dashed left
\draw [ultra thick, dashed] (3.5,5.) -- (3.5,6);
\draw [ultra thick, dashed] (3.5,3)  to [out=45,in=-45] (3.5,5);
\draw [ultra thick, dashed] (3.5,1)  to [out=135,in=-135] (3.5,3);
\draw [ultra thick, dashed] (3.5,0.) -- (3.5,1);
\draw [ultra thick, dashed, path fading=east] (3.5,0) -- (4,0.3);
\draw [ultra thick, dashed, path fading=east] (3.5,6) -- (4,6.3);
\draw [ultra thick, dashed, path fading=east] (3.5,5) -- (4,5.3);
\draw [ultra thick, dashed, path fading=east] (4.,4) -- (4.5,4.3);
\draw [ultra thick, dashed, path fading=east] (3.5,3) -- (4,3.3);
\draw [ultra thick, dashed, path fading=east] (3.1,2) -- (3.6,2.3);
\draw [ultra thick, dashed, path fading=east] (3.5,1) -- (4,1.3);

%top and bottom dashed middle
\draw [ultra thick, dashed] (3.5,5) --(6.5,5);
\draw [ultra thick, dashed, path fading=east] (4.5,5) -- (5,5.3);
\draw [ultra thick, dashed, path fading=east] (5.5,5) -- (6,5.3);
\draw [ultra thick, dashed] (3.5,1) --(6.5,1);
\draw [ultra thick, dashed, path fading=east] (4.5,1) -- (5,1.3);
\draw [ultra thick, dashed, path fading=east] (5.5,1) -- (6,1.3);

%1 filling
\shade[top color=lightgray, bottom color=lightgray, opacity=0.3] (-0.5,1.) to (-0.5,0.) to (3.5,0) to (3.5,1) to (-0.5,1);
\shade[top color=lightgray, bottom color=lightgray, opacity=0.3] (-0.5,1) to (-1,2) to (-0.5,3) to (3.5,3) to [out=-135,in=135] (3.5,1) to (-0.5,1);
\shade[top color=lightgray, bottom color=lightgray, opacity=0.3] (-0.5,5.) to (3.5,5.) to [out=-45,in=45] (3.5,3) to (-0.5,3) to (-1,4) to (-0.5,5);
\shade[top color=lightgray, bottom color=lightgray, opacity=0.3] (-0.5,5.) to (-0.5,6.) to (3.5,6) to (3.5,5) to (-0.5,5);

%2 filling
\shade[top color=lightgray, bottom color=lightgray, opacity=0.9] (6.5,1.) to (6.5,0.) to (10.,0) to (10.,1) to (6.5,1);
\shade[top color=lightgray, bottom color=lightgray, opacity=0.9] (6.5,1) to [out=135,in=-90] (5.75,2.125)
to [out=90,in=-135] (6.5,3.) to (10.,3) to (9.5,2) to (10.,1) to (6.5,1);
\shade[top color=lightgray, bottom color=lightgray, opacity=0.9] (6.5,5.) to (10.,5.) to (9.5,4) to (10.,3) to (6.5,3) to [out=45,in=-90]  (7.25,3.875) to [out=90,in=-45] (6.5,5);
\shade[top color=lightgray, bottom color=lightgray, opacity=0.9] (6.5,5.) to (6.5,6.) to (10.,6) to (10.,5) to (6.5,5);

%\path [fill=lightgray, opacity=0.9] (6.5,1) -- (6.5,5) -- (10,5) -- (9.5,4) -- (10,3) -- (9.5,2) -- (10,1) -- (6.5,1);

% magnetic field arrows middle
\draw [ultra thick, red, -stealth] (4.76,4)  to  (4.75,4.1);
\draw [ultra thick, red, path fading = south] (4.5,3.25)  to [out=45,in=-45]  (4.25,5);
\draw [ultra thick, red, path fading = north] (4.25,1)  to [out=135,in=-135] (4.25,2.75);
\draw [ultra thick, red, -stealth] (3.89,1.8)  to  (3.89,1.9);

\draw [ultra thick, red, -stealth] (5.515,4)  to  (5.505,4.1);
\draw [ultra thick, red, path fading = south] (5.25,3.25)  to [out=45,in=-45]  (5.,5);
\draw [ultra thick, red, path fading = north] (5.,1)  to [out=135,in=-135] (4.75,2.75);
\draw [ultra thick, red, -stealth] (4.52,1.8)  to  (4.50,1.9);

\draw [ultra thick, red, -stealth] (6.415,4)  to  (6.39,4.1);
\draw [ultra thick, red, path fading = south] (6.25,3.45)  to [out=45,in=-45]  (5.75,5);
\draw [ultra thick, red, path fading = north] (5.75,1)  to [out=135,in=-135] (5.25,2.55);
\draw [ultra thick, red, -stealth] (5.145,1.8)  to  (5.11,1.9);

% magnetic field arrows left
\draw [ultra thick, red, -stealth] (2.76,4)  to  (2.75,4.1);
\draw [ultra thick, red, path fading = south] (2.5,3.25)  to [out=45,in=-45]  (2.25,5);
\draw [ultra thick, red, path fading = north] (2.25,1)  to [out=135,in=-135] (2.25,2.75);
\draw [ultra thick, red, -stealth] (1.89,1.8)  to  (1.89,1.9);

\draw [ultra thick, red, -stealth] (1.16,4)  to  (1.15,4.1);
\draw [ultra thick, red, path fading = south] (1,3.25)  to [out=60,in=-60]  (0.75,5);
\draw [ultra thick, red, path fading = north] (0.75,1)  to [out=120,in=-120] (0.75,2.75);
\draw [ultra thick, red, -stealth] (0.5,1.8)  to  (0.5,1.9);

% magnetic field arrows right

\draw [ultra thick, red, -stealth] (7.81,4)  to  (7.79,4.1);
\draw [ultra thick, red, path fading = south] (7.65,3.45)  to [out=45,in=-45]  (7.15,5);
\draw [ultra thick, red, path fading = north] (7.15,1)  to [out=135,in=-135] (6.65,2.55);
\draw [ultra thick, red, -stealth] (6.545,1.8)  to  (6.51,1.9);

\draw [ultra thick, red, -stealth] (8.21,4)  to  (8.19,4.1);
\draw [ultra thick, red, path fading = south] (8.05,3.45)  to [out=50,in=-50]  (7.65,5);
\draw [ultra thick, red, path fading = north] (7.65,1)  to [out=130,in=-130] (7.25,2.55);
\draw [ultra thick, red, -stealth] (7.145,1.8)  to  (7.11,1.9);

\draw [ultra thick, red, -stealth] (8.62,4)  to  (8.6,4.1);
\draw [ultra thick, red, path fading = south] (8.45,3.35)  to [out=55,in=-55]  (8.15,5);
\draw [ultra thick, red, path fading = north] (8.15,1)  to [out=125,in=-125] (7.85,2.65);
\draw [ultra thick, red, -stealth] (7.725,1.8)  to  (7.69,1.9);

\draw [ultra thick, red, -stealth] (9.02,4)  to  (9,4.1);
\draw [ultra thick, red, path fading = south] (8.85,3.25)  to [out=60,in=-60]  (8.65,5);
\draw [ultra thick, red, path fading = north] (8.65,1)  to [out=120,in=-120] (8.45,2.75);
\draw [ultra thick, red, -stealth] (8.31,1.8)  to  (8.29,1.9);

\draw [ultra thick, red, -stealth] (9.42,4)  to  (9.41,4.1);
\draw [ultra thick, red, path fading = south] (9.25,3.25)  to [out=65,in=-65]  (9.15,5);
\draw [ultra thick, red, path fading = north] (9.15,1)  to [out=115,in=-115] (9.,2.75);
\draw [ultra thick, red, -stealth] (8.87,1.8)  to  (8.85,1.9);

%curved dashed right vertical
\draw [ultra thick, dashed] (6.5,5) --(6.5,6);
\draw [ultra thick, dashed] (6.5,3) to [out=45,in=-90]  (7.25,3.875) to [out=90,in=-45]  (6.5,5);
\draw [ultra thick, dashed] (6.5,1) to [out=135,in=-90]  (5.75,2.125) to [out=90,in=-135]  (6.5,3);
\draw [ultra thick, dashed] (6.5,0) --(6.5,1);

%curved dashed right horizontal
\draw [ultra thick, dashed, path fading=east] (6.5,6) -- (7,6.3);
\draw [ultra thick, dashed, path fading=east] (6.5,5) -- (7,5.3);
\draw [ultra thick, dashed, path fading=east] (7.25,4) -- (7.75,4.3);
\draw [ultra thick, dashed, path fading=east] (6.5,3) -- (7,3.3);
\draw [ultra thick, dashed, path fading=east] (5.75,2) -- (6.25,2.3);
\draw [ultra thick, dashed, path fading=east] (6.5,1) -- (7,1.3);
\draw [ultra thick, dashed, path fading=east] (6.5,0) -- (7,0.3);

%bottom and top dashed right
\draw [ultra thick, dashed] (6.5,5) --(10,5);
\draw [ultra thick, dashed, path fading=east] (7.5,5) -- (8,5.3);
\draw [ultra thick, dashed, path fading=east] (8.5,5) -- (9,5.3);
\draw [ultra thick, dashed, path fading=east] (9.5,5) -- (10,5.3);
\draw [ultra thick, dashed] (6.5,1) --(10,1);
\draw [ultra thick, dashed, path fading=east] (7.5,1) -- (8,1.3);
\draw [ultra thick, dashed, path fading=east] (8.5,1) -- (9,1.3);
\draw [ultra thick, dashed, path fading=east] (9.5,1) -- (10,1.3);

\draw [<-] (0,6) -- (0,-0.3); 	%z
\draw [->] (0,1) -- (10,1);		%x
\draw [->] (0,1) -- (0.5,1.3);	%y

\small

\node [below] at (10,1) {$x$};
\node [left] at (0,6) {$z$};
\node [right] at (0.45,1.3) {$y$};

\large
\node [right] at (0.35,3) {$\rho_1$,$p_1$,$T_1$,$B_1$};
\node [right] at (3.75,3) {$\rho_0$,$p_0$,$T_0$,$B_0$};
\node [right] at (7.1,3) {$\rho_2$,$p_2$,$T_2$,$B_2$};
\node [left] at (-0.5,1) {$z=0$};
\node [left] at (-0.5,5) {$z=L$};
\node [below] at (3.5,-0.2) {$x=-x_0$};
\node [below] at (6.5,-0.2) {$x=x_0$};
%magnetic field end points
\fill[red] (4.25,5) circle (0.08cm);
\fill[red] (4.25,1) circle (0.08cm);
\fill[red] (5,5) circle (0.08cm);
\fill[red] (5,1) circle (0.08cm);
\fill[red] (5.75,5) circle (0.08cm);
\fill[red] (5.75,1) circle (0.08cm);
\fill[red] (0.75,5) circle (0.08cm);
\fill[red] (0.75,1) circle (0.08cm);
\fill[red] (2.25,5) circle (0.08cm);
\fill[red] (2.25,1) circle (0.08cm);
\fill[red] (7.15,5) circle (0.08cm);
\fill[red] (7.15,1) circle (0.08cm);
\fill[red] (7.65,5) circle (0.08cm);
\fill[red] (7.65,1) circle (0.08cm);
\fill[red] (8.15,5) circle (0.08cm);
\fill[red] (8.15,1) circle (0.08cm);
\fill[red] (8.65,5) circle (0.08cm);
\fill[red] (8.65,1) circle (0.08cm);
\fill[red] (9.15,5) circle (0.08cm);
\fill[red] (9.15,1) circle (0.08cm);

\end{tikzpicture}
}
\caption{Same as Figure \ref{fig:standing_kink_fund} but for the first harmonic.}
\label{fig:standing_kink_first}
  \end{minipage}
\end{figure}

\section{Amplitudes of Standing Harmonic Modes} \label{Amplitudes of the Standing Harmonic Modes}
The focus in this section is to derive the amplitude difference between the two sides of the magnetic slab embedded in a magnetically asymmetric environment. More precisely, we will take the magnitude of the difference of the amplitudes of the standing oscillations at the two sides of the slab. This quantity will be denoted $D_{S}$ and $D_{K}$ for the quasi-sausage and quasi-kink modes, respectively. Using Equation (\ref{dr5}), the solution for the velocity perturbation inside the slab can be written as $\hat{v}_x=B\cosh(m_0x)+C\sinh(m_0x)$, and the arbitrary constants B and C will appear in the solutions for the amplitude difference between the two sides of the slab. The relation $\hat{\xi}_x(x)=i\hat{v}_x(x)/\omega$ is used at $x=\pm x_0$, where $\hat{\xi}_x(x_0)$ and $\hat{\xi}_x(-x_0)$ are the amplitudes of oscillation at the slab boundaries \citep{all-erd-18}. Similarly, as with the eigenfrequencies, the precise details of the derivation are not included. Counterpart similar quantities are derived in both \cite{ox-20} and \cite{all-erd-18}, and more information regarding the derivation can be found in both of these studies. Care must be taken once again as the solutions are only valid under certain orderings of characteristic speeds as discussed where the eigenfrequencies are presented in Section \ref{Frequencies of the Standing Harmonic Modes}.

The amplitude difference between the two sides of the magnetic slab in a magnetically asymmetric environment takes the following from for both the quasi-sausage and quasi-kink oscillations:

\begin{equation}
\abs{\abs{\hat{\xi}_x(x_0)}-\abs{\hat{\xi}_x(-x_0)}} = \abs{\frac{1}{\omega}\left(\abs{\hat{v}_x(x_0)}-\abs{\hat{v}_x(-x_0)}\right)}.
\label{as}
\end{equation}

\subsection{Quasi-sausage Modes} \label{QS2}

In the case of the quasi-sausage modes, $\hat{v}_x(x_0)$ and $\hat{v}_x(-x_0)$ (along with $\hat{\xi}_x(x_0)$ and $\hat{\xi}_x(-x_0)$) will have opposite signs, so Equation (\ref{as}) reduces to

\begin{equation}
D_{S}= \abs{\hat{\xi}_x(x_0)+\hat{\xi}_x(-x_0)} = \abs{\frac{1}{\omega}\left(\hat{v}_x(x_0)+\hat{v}_x(-x_0)\right)}.
\label{as1}
\end{equation}
The amplitude difference of the quasi-sausage mode corresponding to the eigenfrequency given by Equation (\ref{fi2}) is

\begin{multline}
D_{S} \approx \zeta\sqrt{\varepsilon}\abs{C}\frac{\sqrt{-\bar{\Pi}}L\rho_0v_{A0}^2(c_0^2+v_{A0}^2)^{1/2}(c_1^2-c_{T0}^2)^{1/2}}{\sqrt{n\pi}\rho_1c_0^2(c_{T0}^2-v_{A1}^2)^{1/2}(c_{T0}^2-c_{T1}^2)^{1/2}(c_1^2+v_{A1}^2)^{1/2}}\times \\ \abs{\frac{v_{A1}^2}{2(v_{A1}^2-c_{T0}^2)}\left(1+\frac{(c_1^2-c_{T0}^2)(v_{A1}^2-c_{T0}^2)}{(c_{T1}^2-c_{T0}^2)(c_1^2+v_{A1}^2)}\right)-\frac{v_{A1}^2\gamma}{2c_1^2+v_{A1}^2\gamma}},
\label{as2}
\end{multline}
where the ordering $v_{A1}<c_{T0}<c_1$ is taken, just as with the solution for the eigenfrequency given in Equation (\ref{fi2}).

Next, the amplitude difference of the quasi-sausage mode corresponding to the eigenfrequency given by Equation (\ref{fi6}) is

\begin{equation}
D_{S} \approx \zeta\varepsilon \abs{C} \frac{\rho_0L\sqrt{\abs{\tilde{\Pi}}}\abs{v_{A0}^2-c_1^2}^{1/2}\abs{c_{T0}^2-c_1^2}^{1/2}(c_0^2+v_{A0}^2)^{1/2}}{\rho_1c_1^2\abs{c_0^2-c_1^2}^{1/2}\abs{v_{A1}^2-c_1^2}^{1/2}}\abs{\frac{v_{A1}^2}{2(v_{A1}^2-c_1^2)}-\frac{\gamma v_{A1}^2}{2c_1^2+\gamma v_{A1}^2}},
\label{as3}
\end{equation}
where, for both solutions, $C$ is an arbitrary constant that can be set equal to one. The amplitude difference of oscillations between the two sides of the magnetic slab, for both quasi-sausage frequencies, shows a rather complex dependence on the characteristic speeds. Both expressions are linearly dependent on the magnetic asymmetry parameter $\zeta$, while both solutions are directly related to $\varepsilon$ through different exponents. The first amplitude difference, given by Equation (\ref{as2}), shows a square-root dependence on $\varepsilon$, while the second one, given by Equation (\ref{as3}), simply shows a linear relationship with $\varepsilon$. Due to the assumptions that $m_0^2,m_1^2>0$, and the consequently ordered characteristic speeds, the first solution, given by Equation (\ref{as2}), has been simplified.

A comparison to the counterpart quantities given in \cite{ox-20} can be noted here, however, the solutions given by Equations (\ref{as2}) and (\ref{as3}) both contain a factor of the magnetic asymmetry parameter $\zeta$, and as this represents a different type of asymmetry to that considered in \cite{ox-20}, the equations here do not reduce to any in their study.

\subsection{Quasi-kink Modes} \label{QK2}

Even though a solution for the quasi-kink frequencies was not determined with the inclusion of a term representing magnetic asymmetry explicitly, we can still calculate the amplitude difference between the two sides of the magnetic slab to leading-order. This is because, we only need to use a solution for eigenfrequency that includes the first correction term (i.e. the term of order $\varepsilon^2$). These solutions are given in Equations (\ref{fi10}) and (\ref{fi12}).

In the case of the quasi-kink modes, $\hat{v}_x(x_0)$ and $\hat{v}_x(-x_0)$ (along with $\hat{\xi}_x(x_0)$ and $\hat{\xi}_x(-x_0)$) will have the same sign, so Equation (\ref{as}) reduces to

\begin{equation}
D_{K}= \abs{\hat{\xi}_x(x_0)-\hat{\xi}_x(-x_0)} = \abs{\frac{1}{\omega}\left(\hat{v}_x(x_0)-\hat{v}_x(-x_0)\right)}.
\label{ak1}
\end{equation}
The quasi-kink amplitude difference corresponding to the eigenfrequency given by Equation (\ref{fi10}) is

\begin{equation}
D_{K} \approx \Bigg|\zeta B \frac{Lv_{k}^5\rho_1^2(c_0^2-v_k^2)}{2n\pi\rho_0^2(c_{T0}^2-v_k^2)(c_0^2+v_{A0}^2)(v_{A0}^2-v_k^2)(v_k^2-c_1^2)}\Bigg|.
\label{ak2}
\end{equation}
The quasi-kink amplitude difference corresponding to the eigenfrequency given by Equation (\ref{fi12}) is

\begin{equation}
D_{K} \approx \zeta \abs{B} \frac{Lv_{A1}^2c_{TA}^3\rho_1^2(c_0^2-c_{TA}^2)}{2n\pi\rho_0^2(c_{TA}^2-c_{T0}^2)(c_0^2+v_{A0}^2)(c_{TA}^2-v_{A0}^2)(c_1^2-c_{TA}^2)},
\label{ak3}
\end{equation}
where the ordering $v_{A0}<c_{TA}<c_0$ is taken, just as with the solution for the frequency given in Equation (\ref{fi12}).

For both quasi-kink amplitude difference solutions $B$ is an arbitrary constant that can be set equal to one. The amplitude difference between the two sides of the slab, for both quasi-kink eigenfrequencies, also shows a complex dependence on the characteristic speeds, just as for the quasi-sausage modes. Both are linearly dependent on the magnetic asymmetry parameter $\zeta$, and do not depend on $\varepsilon$ to leading-order. Due to the assumptions that $m_0^2,m_1^2>0$, and the consequently ordered characteristic speeds, the second solution, given by Equation (\ref{ak3}), has been simplified.

As discussed in Section \ref{QK1}, none of the derived frequencies for the quasi-kink modes can be compared to those derived in \cite{ox-20}, and consequently the above obtained amplitude expressions cannot be compared with those derived by \cite{ox-20}.

\section{Discussion}
This study has built on the MHD wave studies in symmetric magnetic slab models introduced by \cite{rob-82}, which was further generalised for studying propagating MHD waves by \cite{zsam-all-18} through the inclusion of magnetic asymmetry, by considering now standing waves. Our work, here, is also a generalisation of \cite{ox-20} who considered the asymmetry in external density with no external magnetic fields present. Now, we introduced asymmetry between the external magnetic fields, and examined its effect on the standing modes. For analytical progress, the thin slab and weak magnetic asymmetry approximations are used.

The derived equations for the eigenfrequencies of the standing harmonic modes in Section \ref{Frequencies of the Standing Harmonic Modes} show that they are more prone to changes in the width of the magnetic slab than they are to changes in the external magnetic field, under the assumption of an isothermal plasma.

Where it was possible to derive equations for the eigenfrequency which included dependence on $\zeta$, which happened to be the quasi-sausage modes, the frequency ratio of the first harmonic to the fundamental mode was also calculated. Such frequency ratios are popular tools when conducting the analysis of signatures of MHD wave observations with diagnostics in mind. It is confirmed that as the parameter $\varepsilon$ is reduced to zero, so the slab becomes infinitely thin, the frequency ratio is reduced to $\omega_2/\omega_1=2$. This is as expected as the slab should then behave like a homogeneous magnetic string with fixed ends, representing line-tying. Note that as we take $\varepsilon \rightarrow 0$, we also need to reduce the \value of $\zeta$ to obey the ordering used to derive the analytic expressions. However, provided $\zeta$ is non-zero, there is still a tangential discontinuity (i.e. magnetic interface) and surface waves may exist.

The equations derived for the eigenfrequency ratio of the first harmonic to the fundamental mode can be applied to observational results. Both solutions, given by Equations (\ref{fi3}) and (\ref{fi7}), could be inverted and an expression for the asymmetry parameter $\zeta$ could be determined. Then, given an observational measurement for the frequency ratio $\omega_2/\omega_1$, it would be possible to calculate the percentage difference between the magnetic field strength on the two sides of the slab, under the assumption that e.g. density remains constant and the asymmetry comes only from the magnetic fields. Of course, if density-sensitive intensity measurements would accompany the frequency ratio measurements, the restriction on assuming a constant density may be removed. Here, for the sake of making our point, we retain the assumption on density.

Consider, as an example, a sunspot light bridge using the magnetic slab model presented in this study, with approximate estimates of width and length given by 0.7 Mm and 2.5 Mm \citep{sch-16, yang-17}, respectively, giving $\varepsilon \approx 0.14$. Estimates of internal and external sound speeds are $c_0 \approx 7$ km/s and $c_1 \approx 6$ km/s \citep{sob-13}. Estimating the plasma-$\beta$ in the interior and on the left-hand side, say, $\beta_0 \approx \beta_1 \approx 2.5$ \citep{bor-11, fel-16, liu-16}, allows approximate values for the Alfv\'en speed to be calculated, and consequently approximate values for the tube speed can be obtained. Using pressure balance given by Equation (\ref{slab2}), a value for the ratio of the external (recall, here we use $\rho_1=\rho_2$) to the internal density can be determined (taking $\gamma=5/3$). Now, assume a deviation of say $2.5\%$ is observed in the frequency ratio, $\omega_2/\omega_1\approx 1.95$. Using Equation (\ref{fi7}), an estimate for $\zeta$ can now be determined. Consequently, a value for the percentage difference between the magnetic field strength on the two sides of the slab can be calculated. Applying the explicit values given here, it is then found that the magnetic field strength on the right-hand side is $\approx 8.5\%$ larger than the left-hand side.

The amplitude difference between the two sides of the magnetic slab, a potentially practical quantity to be utilised through solar spatio-magneto-seismology for e.g. MHD diagnostics, was also calculated for both the quasi-sausage and quasi-kink oscillations. The amplitude differences reduce to zero when the magnetic asymmetry is removed from the system, as expected. The amplitude difference is a potentially observable quantity, and combining observational data with inversion of the equation for the amplitude difference could yield expressions for the magnetic field strength in one of the regions. Applying such inversions to the observations provides an example of the power of solar magneto-seismology. It is expected that observations with ultra-high spatial resolution will be needed that only may be yielded by the capabilities of Daniel K. Inouye Solar Telescope (DKIST). 

In \cite{ox-20}, a similar procedure was employed to study a magnetic slab embedded in a non-magnetic asymmetric environment, and the eigenfrequencies and amplitudes were analysed under the thin slab and weak asymmetry assumptions. As the present study provides a similar analysis with the complication of an external magnetic field, a comparison can be carried out between the solutions for eigenfrequencies and amplitudes of the two studies. It is evident that the eigenfrequency given by Equation (\ref{fi2}) reduces to that given by Equation (26) in \cite{ox-20} provided the asymmetry in both studies is reduced to zero and the external magnetism is removed from this model. Due to the different types of asymmetry considered in the two studies, terms involving asymmetry are not compared. Further, there are solutions presented here that have no counterpart in the magnetic slab embedded in a non-magnetic external environment.

A derivation of estimating both the relative eigenfrequency and the relative amplitude differences due to asymmetry were carried out in \cite{ox-20}. By considering the phenomenon of weak asymmetry as a perturbation to a symmetric magnetized plasma slab, an application of the Rayleigh-Ritz technique was made. Here, we have carried out a derivation of the eigenfrequencies of the quasi-sausage waves, under the assumptions of $\zeta$ and $\varepsilon$ being of the same order, and taking a similar approach to that in \cite{ox-20} is possible here. Consider the solution given by Equation (\ref{fi2}), the results would yield that the relative differences in the eigenfrequency due to magnetic asymmetry would be of the order $\zeta \varepsilon$, whereas the relative amplitude difference due to asymmetry would be of the order $\zeta$. This finding shows, as predicted by the Rayleigh-Ritz technique, that the eigenfrequencies are affected by asymmetry to higher-order in small quantities than the amplitudes.

\acknowledgments
R.E. is grateful to Science and Technology Facilities Council (STFC, grant number ST/M000826/1) UK and the Royal Society for enabling this research. R.E. also acknowledges the support received by the CAS Presidents International Fellowship Initiative Grant No. 2019VMA052 and the warm hospitality received at USTC of CAS, Hefei, where part of his contribution was made. N.Z. is also grateful to the School of Mathematics and Statistics at the University of Sheffield, and to the University of Debrecen.

%bibliography
\bibliographystyle{aasjournal}
\bibliography{standing_waves_bibliography}

\end{document}